\documentclass{article}

\usepackage{arxiv}

\usepackage[utf8]{inputenc} % allow utf-8 input
\usepackage[T1]{fontenc}    % use 8-bit T1 fonts
\usepackage{hyperref}       % hyperlinks
\usepackage{url}            % simple URL typesetting
\usepackage{booktabs}       % professional-quality tables
\usepackage{amsfonts}       % blackboard math symbols
\usepackage{nicefrac}       % compact symbols for 1/2, etc.
\usepackage{microtype}      % microtypography
\usepackage{lipsum}
\usepackage{graphicx}

\usepackage{subcaption}

\usepackage{enumitem}

\newtheorem{definition}{Definition}
\title{Understanding Individual-Space Relationships to Inform and Enhance Location-Based Applications}

\author{
 Licia Amichi \\
 Oak Ridge National Laboratory\\
  \texttt{amichil@ornl.gov} \\
   \And
 Gautam Malviya Thakur \\
 Oak Ridge National Laboratory\\
\texttt{thakurg@ornl.gov} \\
  \And
 Carter Christopher \\
  Oak Ridge National Laboratory\\
  \texttt{christophesc@ornl.gov} \\
}

\begin{document}
\maketitle
\begin{abstract}
Understanding the complex dynamics of human navigation and spatial behavior is essential for advancing location-based services, public health, and related fields. This paper investigates the multifaceted relationship between individuals and their environments (e.g. location and places they visit), acknowledging the distinct influences of personal preferences, experiences, and social connections. While certain locations hold sentimental value and are frequently visited, others function as mere transitory points. To the best of our knowledge, this paper is the first to exploit visitation patterns and dwell times to characterize an individual's relationship with specific locations. We identify seven key types of spatial relationships and analyze the discrepancies among these visit types across semantic, spatial, and temporal dimensions. Our analysis highlights key findings, such as the prevalence of anchored-like visits (e.g. home, work) in both real-world Singapore and Beijing datasets, with unique associations in each city -Singapore's anchored-liked visits include recreational spaces, while Beijing's are limited to residential, business, and educational sites. These findings emphasize the importance of geographic and cultural context in shaping mobility and their potential in benefiting the precision and personalization of location-based services.
\end{abstract}

% keywords can be removed

\section{Introduction}
In today's world, people have the opportunity to travel to many different places 
throughout their lifetimes and in short spans. Every day, millions of people spend billions of hours collectively on their commutes. According to a recent survey by Agoda, 70\% of the world population have visited up to 10 countries~\cite{agoda_2019}. For instance, 2022 scored over 963 million international tourists~\cite{unwto_2023}. Mobility is an essential aspect of our daily lives, and with urbanization, globalization, and advancements in transportation, it is expected to continue to grow~\cite{barbosa2018human,pappalardo2023future,10.1145/3240323.3240361}.

Despite the abundance of data generated by modern transportation systems, location-based services, and digital footprints, a comprehensive understanding of the interplay between individuals and their surroundings remains elusive~\cite{10.1145/3494993,cuttone2018understanding}. Questions persist regarding the determinants of location preference, variations in mobility patterns, and the delineation of routine locations. While numerous studies have explored human mobility through various lenses, existing approaches often fall short due to oversimplified mechanisms~\cite{10.1145/3240323.3240361,Huandong_2022,10.1145/3494993}. The literature has witnessed significant innovation in mobility modeling, spanning from probabilistic models to deep learning and generative AI approaches~\cite{pappalardo2024survey,10.1145/3494993}. Traditional mechanistic models assume that human mobility behavior obeys simplistic governing laws such as heavy-tailed displacements and waiting times, a slow-down tendency for location discovery, or preferential return~\cite{gonzalez2008understanding,song2010modelling}. However, these conventional methods often fall short of capturing the full spectrum of human mobility behavior. AI techniques like Convolutional Neural Networks (CNNs) and Generative Adversarial Networks (GANs) have emerged to address these limitations. CNNs can incorporate external contextual factors, while GANs excel in capturing diverse and nonlinear relationships concurrently— aspects that conventional models may overlook. Nonetheless, these AI models lack transparency, obscuring the rationale behind predictions and trajectory generation~\cite{luca2021survey}. 

Recent research explores innovative methodologies rooted in knowledge-driven paradigms. Knowledge Graphs (KG) are a powerful representation framework that encodes structured information about entities and their relationships and have emerged as a promising tool for modeling human mobility~\cite{10.1145/3240323.3240361,Huandong_2022,10.1145/3494993}. By encapsulating mobility behaviors as relational facts within a structured KG and leveraging advanced Knowledge Graph Embedding (KGE) techniques, authors in~\cite{10.1145/3677019} aim to capture the complexity of individual mobility patterns. However, applying KG to model the spatial-temporal mobility behavior of users poses several key challenges. Different individuals demonstrate diverse mobility patterns influenced by various factors such as occupation, lifestyle, and personal preferences. A student's mobility patterns may differ significantly from those of a working professional. Capturing and representing this diversity within a single KG framework is a significant challenge due to the complexity and heterogeneity of human behavior and corresponding activities. Furthermore, human mobility behavior often varies over time, with individuals exhibiting different movement patterns on different days or during different times of the day\cite{alessandretti2020scales,pappalardo2024survey}. 

Motivated by the pressing need for a comprehensive understanding of human mobility dynamics, this paper first attempts to explain the complex relationship between individuals and their environments. Specifically, we propose to characterize an individual's connection to a particular place based on the time spent there and the frequency of visits to that place. Through extensive evaluations and comparative analyses, we demonstrate the effectiveness of our proposed methodology in characterizing individual visits, underscoring its practical significance for location-based services, public health interventions, and transportation management strategies.
In summary, our paper makes the following contributions:

\begin{itemize}[leftmargin=*]
    \item \textbf{Data completeness:} We introduce two novel metrics, \textit{temporal completeness} and \textit{spatial completeness}, designed to evaluate the quality and continuity of mobility data. These metrics enable the identification of users with consistently recorded trajectories over extended periods, ensuring more reliable datasets for mobility analysis.

    \item \textbf{Characterization of individual relationships with spatial instances:} We introduce a novel method for characterizing individuals' interactions with spatial locations, incorporating visit frequency and dwell time to classify different types of visits. This allows us to distinguish between routine, exploratory, and casual interactions with locations, shedding light on the complex ways individuals engage with their environments. Our method provides a nuanced understanding of mobility behaviors, revealing patterns such as the transition from exploratory to anchor locations or the persistence of casual visits.

    \item \textbf{Visits characterization evaluation:} We thoroughly evaluate the proposed method using real-world data from Singapore and Beijing, analyzing visit types across semantic, temporal, and spatial dimensions. Our results show that the classification framework effectively distinguishes between predictable mobility patterns and those requiring further investigation. We find that Singapore exhibits a fluid mix of routine and exploratory behaviors, while Beijing shows a stronger adherence to key locations. Semantic categories further highlight city-specific patterns, emphasizing the need for mobility models to reflect geographic and cultural contexts. These insights can enhance recommendation systems by better adapting to diverse mobility behaviors and preferences.
\end{itemize}

\section{Mobility Data Processing}
In this section, we begin by presenting the datasets used in our study, highlighting their key characteristics. We then describe the pre-processing procedures applied to the data, which ensure its suitability for the subsequent analyses tasks.

\subsection{Datasets}
In our experiments, we use real-world mobility datasets and Points of Interest (PoIs) data from two Areas of Interest (AOIs), Singapore and Beijing, to assess the performance of our proposed framework.

\vspace{.1cm}
\noindent\textbf{Singapore dataset:}
The Singapore dataset comprises an extensive collection of daily mobility records for 144,795 users, spanning a two-month period from December 1, 2022, to January 31, 2023. Data points were sampled at intervals of the order of a few seconds. With a total of 264,246,458 data records, this dataset provides a comprehensive and granular view of user mobility patterns.

\vspace{.1cm}
\noindent\textbf{Geolife dataset:}
The Geolife dataset consists of mobility data from 182 users, collected over a period of more than five years, from April 2007 to August 2012. Data points were sampled at intervals of 1 to 5 seconds. These trajectories span over 30 cities, primarily in China, with a concentration in Beijing, along with some data from cities in the USA and Europe, offering a diverse range of mobility patterns across different urban environments.

The Singapore dataset is anonymized, and the data-collecting company adheres to international privacy standards, including the EU General Data Protection Regulation (GDPR) and the California Consumer Privacy Act (CCPA). Similarly, the Geolife dataset follows stringent privacy protocols. Collectively, these measures ensure compliance with established ethical handling standards.

\vspace{.1cm}
\noindent\textbf{PlanetSense PoI dataset:}
We also use the PlanetSense PoI dataset to complement the mobility data by providing contextual information about the locations visited by users. This dataset covers two locations and includes 238,690 PoIs in Singapore and 1,677,835 PoIs in Beijing, categorized into 44 distinct semantic types~\cite{10.1145/3356991.3365474,osti_2000381, thakur2015planetsense}.

\vspace{.1cm}
We begin by introducing the fundamental concepts of human mobility. Let $\mathcal{U}$ and $\mathcal{L}$ denote the sets of $N$ users and $M$ locations, respectively.

\begin{figure*}[ht]
    \centering
    \begin{subfigure}[b]{0.24\textwidth} % Reduced width for fitting all figures in one line
        \centering
        \includegraphics[width=\textwidth]{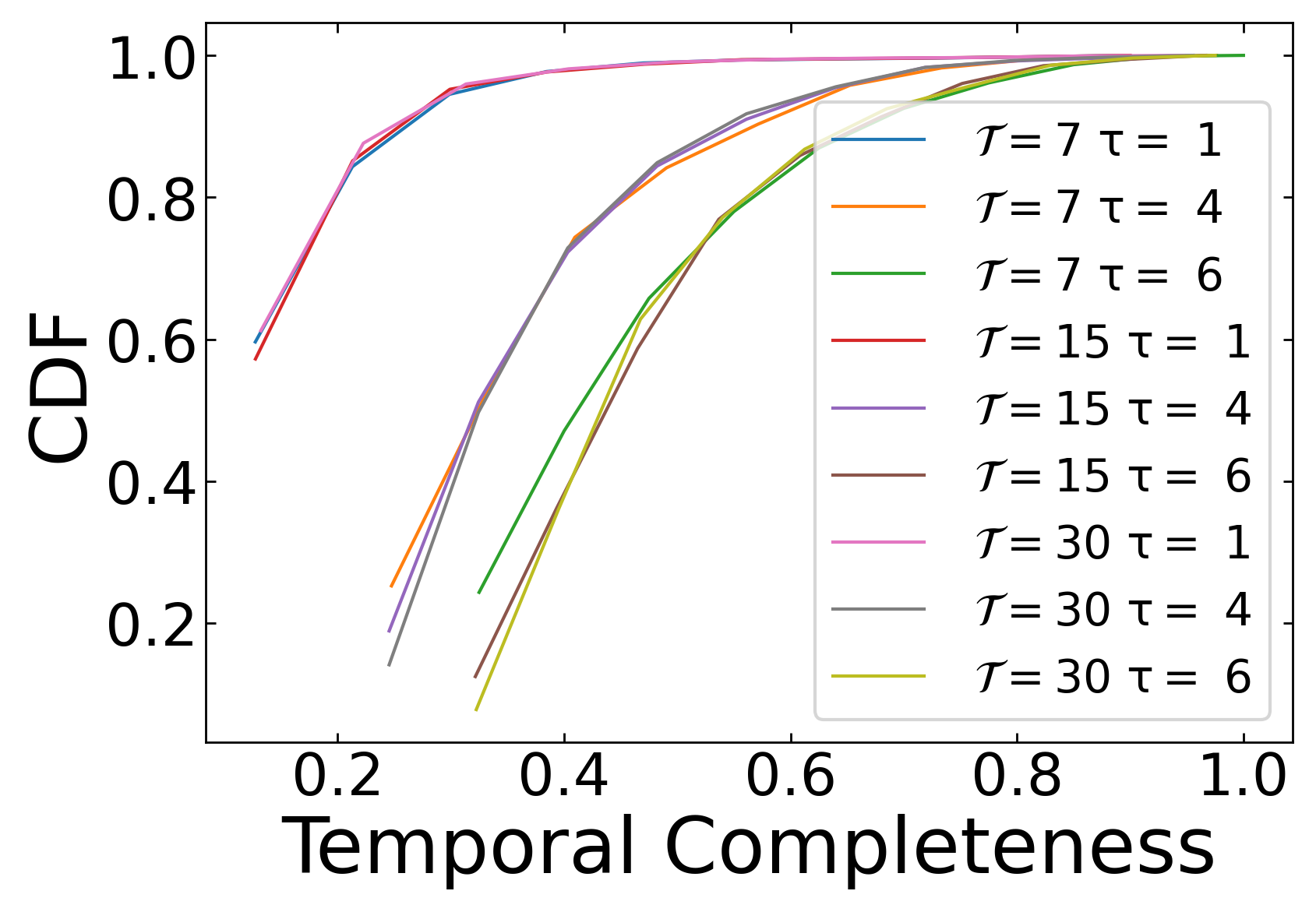}
        \caption{Singapore (temporal).}
        \label{fig:temporal_completeness_singapore}
    \end{subfigure}%
    \hspace{\fill} % Add flexible space between figures
    \begin{subfigure}[b]{0.24\textwidth}
        \centering
        \includegraphics[width=\textwidth]{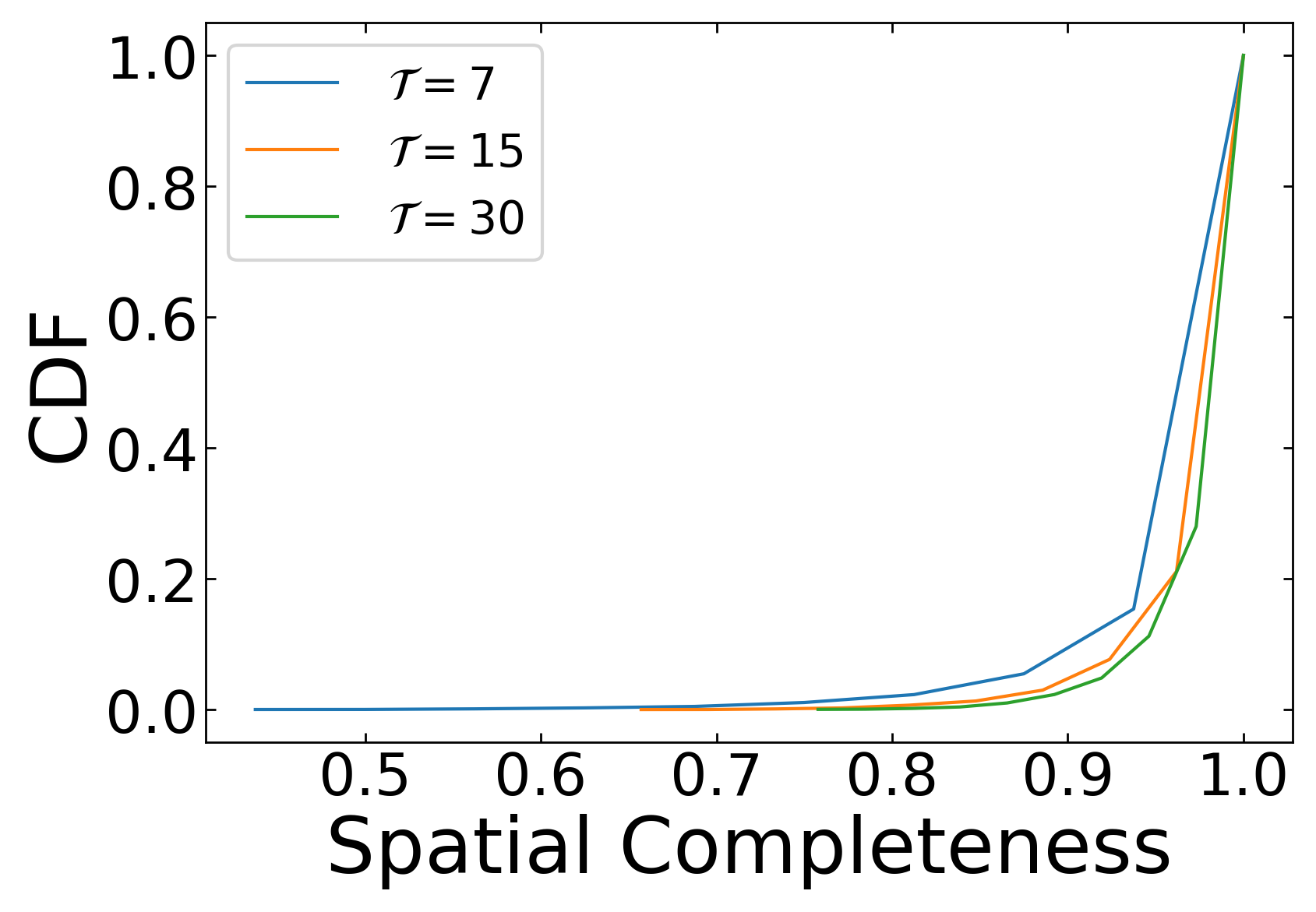}
        \caption{Singapore (spatial).}
        \label{fig:spatial_completeness_singapore}
    \end{subfigure}%
    \hspace{\fill} % Add flexible space between figures
    \begin{subfigure}[b]{0.24\textwidth}
        \centering
        \includegraphics[width=\textwidth]{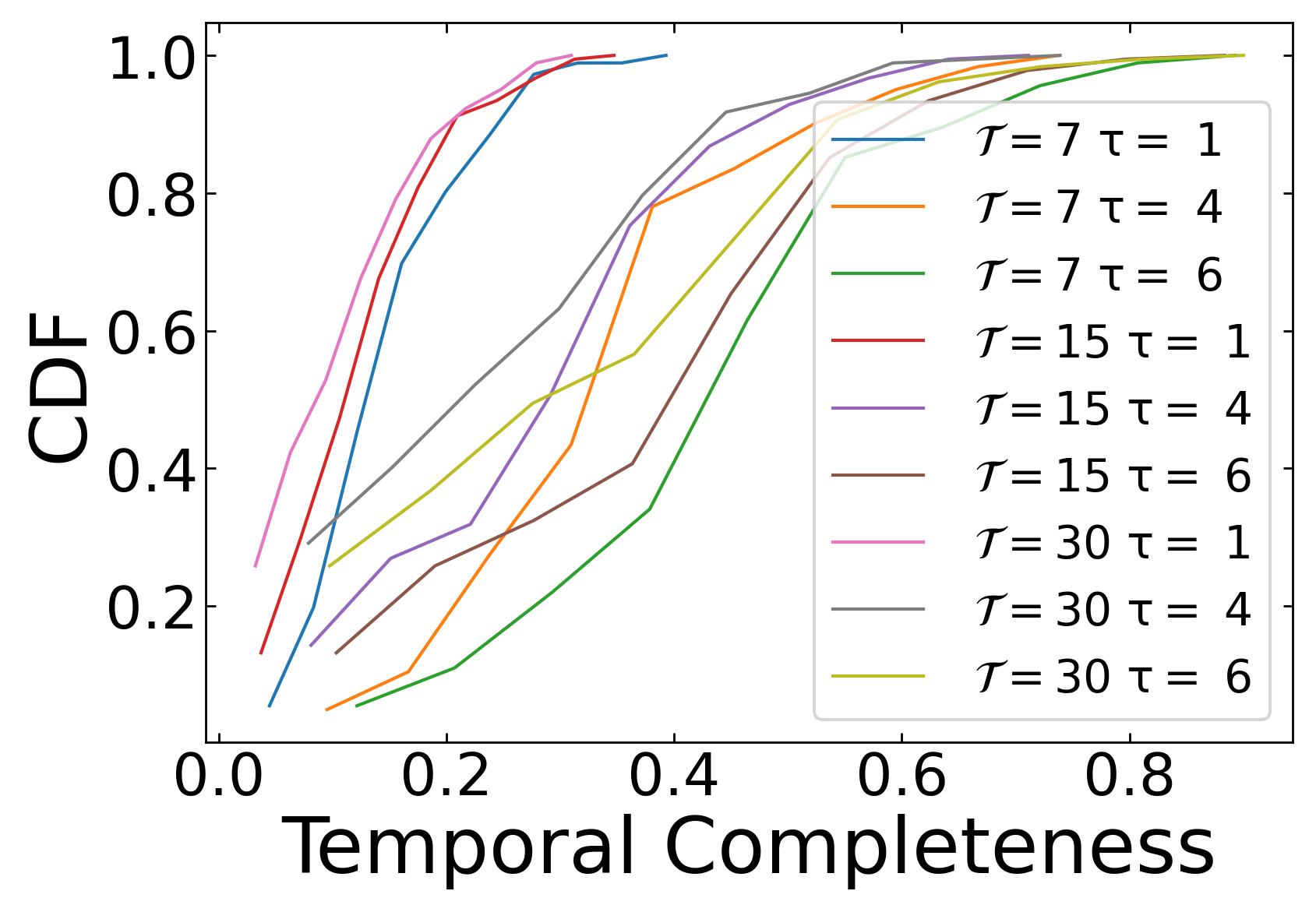}
        \caption{Beijing (temporal).}
        \label{fig:temporal_completeness_beijing}
    \end{subfigure}%
    \hspace{\fill} % Add flexible space between figures
    \begin{subfigure}[b]{0.24\textwidth}
        \centering
        \includegraphics[width=\textwidth]{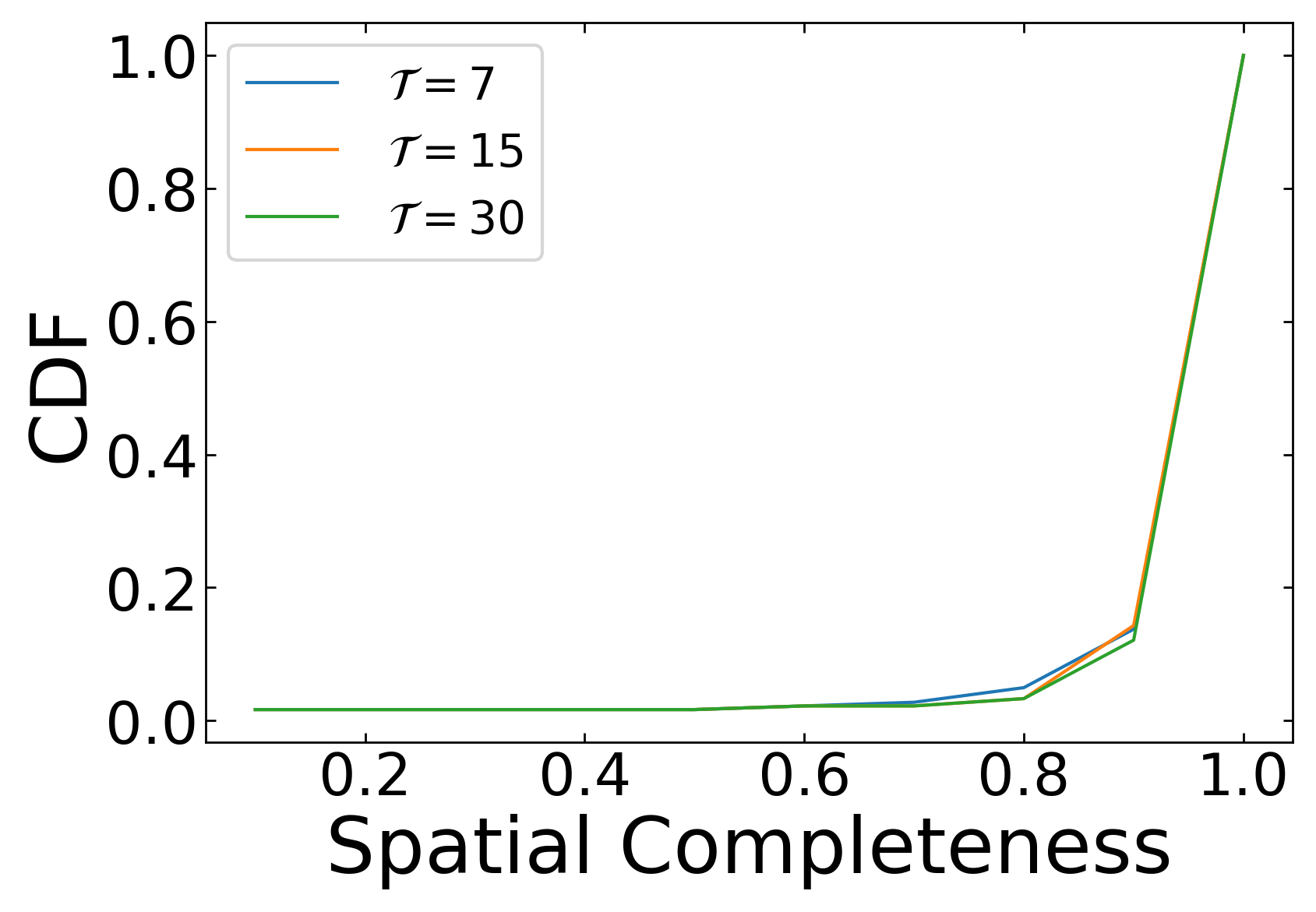}
        \caption{Beijing (spatial).}
        \label{fig:spatial_completeness_beijing}
    \end{subfigure}
    \caption{Temporal and Spatial Completeness in Singapore and Beijing.}
    \label{fig:completeness}
\end{figure*}

\begin{definition}[Mobility Record] 
A mobility record \( m_i \) is defined as a tuple \( m_i = (lat_i, lon_i, t_i) \), where \( (lat_i, lon_i) \) denotes the geographical coordinates of the location visited at time \( t_i \).
\end{definition}

\begin{definition}[Mobility Trace] 
The mobility trace \(\mathcal{D}_u\) of a user \(u\) is a sequence of records \(\{m_i\}_{i=1}^n\). The sequence is ordered such that \(t_1 < t_2 < \ldots < t_n\), ensuring that the tuples are arranged in chronological order.
\end{definition}

\subsection{Data quality assessment metrics}
\subsubsection{Temporal completeness}

Let $\mathcal{D}_u$ represent the mobility trajectory of user $u$ with a total of $N$ observations, recorded over a period $\mathcal{T}$. We define $\tau$ as the window interval size and $P$ as the observation unit. We introduce \textit{temporal completeness} metric $\mu_T(\mathcal{D}_u)$ of a mobility trajectory $\mathcal{D}_u$, as follows,
\begin{equation}
    \mu_T(\mathcal{D}_u) = \frac{\tau}{P} \sum\limits_{i = 0}^{\frac{P}{\tau}} f_{\tau}(i),
\end{equation}

where $f_{\tau}(i) = \left\{
    \begin{array}{ll}
        1 & \mbox{if  } \exists t \in ]\tau \times (i-1), \tau \times i] \mbox{, s.t. } t \in \mathcal{D}_u \\
        0 & \mbox{else.}
    \end{array}
\right.
$

This metric assesses whether mobility records are observed within selected time windows, allowing us to select users with comprehensive temporal mobility patterns.

\subsubsection{Spatial completeness}
Let $\Delta r_i$ represent the distance traveled between locations $l_{i-1}$ and $l_i$, which were visited by the user at times $t_{i-1}$ and $t_i$, respectively. Additionally, let $\Delta t_i$ denote the time elapsed between these two visits. We introduce the \textit{spatial completeness} metric, denoted as $\mu_S(\mathcal{D}_u)$, to quantify the spatial coverage of a mobility trajectory $\mathcal{D}_u$. This metric is defined as follows:
\begin{equation}
    \mu_S(\mathcal{D}_u) = \frac{1}{N} \sum\limits_{i = 1}^{N} g(i),
\end{equation}

where $g(i) = \left\{
    \begin{array}{ll}
        1 & \mbox{if  } \Delta t_i <= P \mbox{ and  } \frac{\Delta r_i}{\Delta t_i} <= \mbox{MAX\_SPEED}\\
        0 & \mbox{else.}
    \end{array}
\right.
$

$MAX\_SPEED$ denotes the maximum plausible speed, serving as a threshold to filter out high speeds that could suggest data errors or anomalies. The \textit{spatial completeness} score ensures that the mobility data accurately reflects realistic movement patterns, avoiding sudden, unrealistic jumps in location.

\subsubsection{Quality assessment}
In this work, we are interested in measuring completeness across various temporal and spatial granularities. To capture daily behaviors effectively, we set the observation unit $P$ to 24 hours, which corresponds to a full day. To evaluate completeness over different time scales, we vary the data availability period $\mathcal{T}$ among one week, two weeks, and one month, i.e., $\mathcal{T} \in \{7, 15, 30\}$ days. These periods are chosen to assess how completeness measures perform over short-term and extended temporal spans. A one-week period helps capture weekly patterns, while two weeks offers insight into possible bi-weekly variations, and a month-long period allows for the evaluation of longer-term trends. Additionally, we vary the window size $\tau$ across $\{1, 4, 6\}$ hours to measure completeness at different temporal granularities. A 1-hour window provides high resolution, enabling detailed analysis of short-term mobility. A 4-hour window balances resolution and data volume, while a 6-hour window captures broader patterns, suitable for identifying less frequent but significant trends. We also set $MAX\_SPEED$ to 150 km/h to filter out movement anomalies, as this value exceeds the typical speed limits of up to 100 km/h in Singapore\footnote{https://www.lta.gov.sg/content/ltagov/en.html} and 120 km/h  in Beijing\footnote{http://www.china.org.cn/bjzt}, thus capturing realistic travel scenarios.

Figure~\ref{fig:completeness} shows that higher temporal granularity (i.e., smaller values of $\tau$) corresponds to lower temporal completeness. This is because a smaller window $\tau$ captures data at a finer time scale, which often results in gaps and reduced completeness. Conversely, increasing the window size $\tau$ typically leads to more comprehensive data coverage for the user.

For the Singapore dataset (see Figure~\ref{fig:temporal_completeness_singapore}), which is richer and denser, the temporal completeness does not show significant variation with changes in the observation period $\mathcal{T}$. This indicates that extending the observation period does not adversely affect data quality. Therefore, we can select a maximum observation period of 30 days without compromising data integrity.

In contrast, the Beijing dataset shows a reduction in temporal completeness with longer observation periods $\mathcal{T}$, as shown in Figure~\ref{fig:temporal_completeness_beijing}. This suggests a trade-off between the desired observation period and data quality. For Beijing, we therefore opt to set the observation period $\mathcal{T}$ to 15 days to balance data quality and coverage.

From Figure~\ref{fig:spatial_completeness_singapore} and~\ref{fig:spatial_completeness_beijing}, the spatial completeness of the data remains consistently high across both AOIs, even when varying the parameters such as window size and observation period. This high spatial completeness suggests that the mobility trajectories captured in the data are reliable and realistic, with minimal occurrences of artifacts such as teleportation or unrealistic jumps in location.

To ensure high data quality in both datasets, we set the observation period $\mathcal{T}$ to 30 days for the Singapore dataset and 15 days for the Beijing dataset. Additionally, we chose a window size $\tau$ of 1 hour for both datasets to achieve a high level of temporal granularity. Subsequently, we filtered and selected only users who have data spanning $\mathcal{T}$ days with a temporal frequency of $\tau = 1$ hour.

\section{Proposed Methodology}

\begin{figure*}[ht]
    \centering
    \begin{subfigure}[b]{0.48\textwidth}
        \centering
        \includegraphics[width=\textwidth]{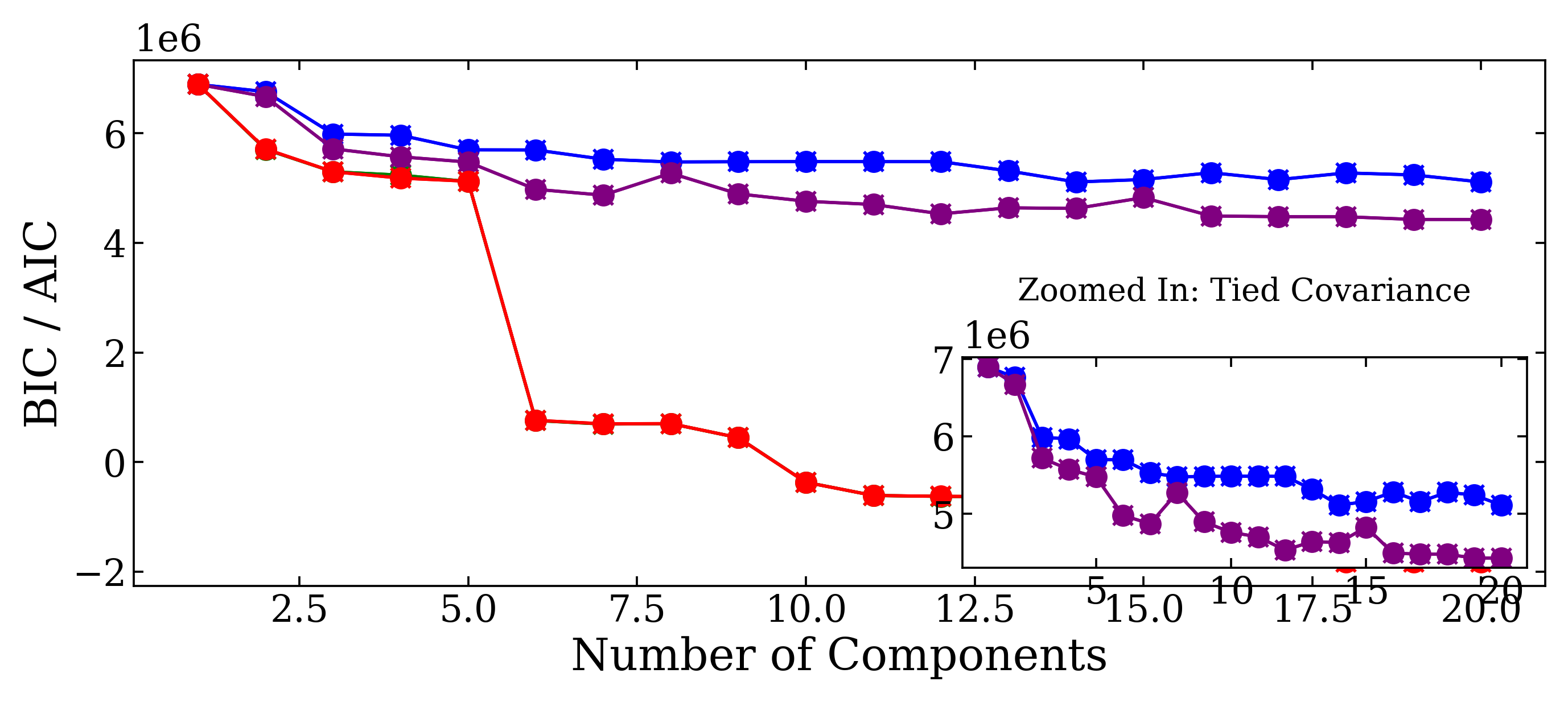}
        \caption{Singapore.}
        \label{fig:bic_aic_singapore}
    \end{subfigure}
    \hfill
    \begin{subfigure}[b]{0.48\textwidth}
        \centering
        \includegraphics[width=\textwidth]{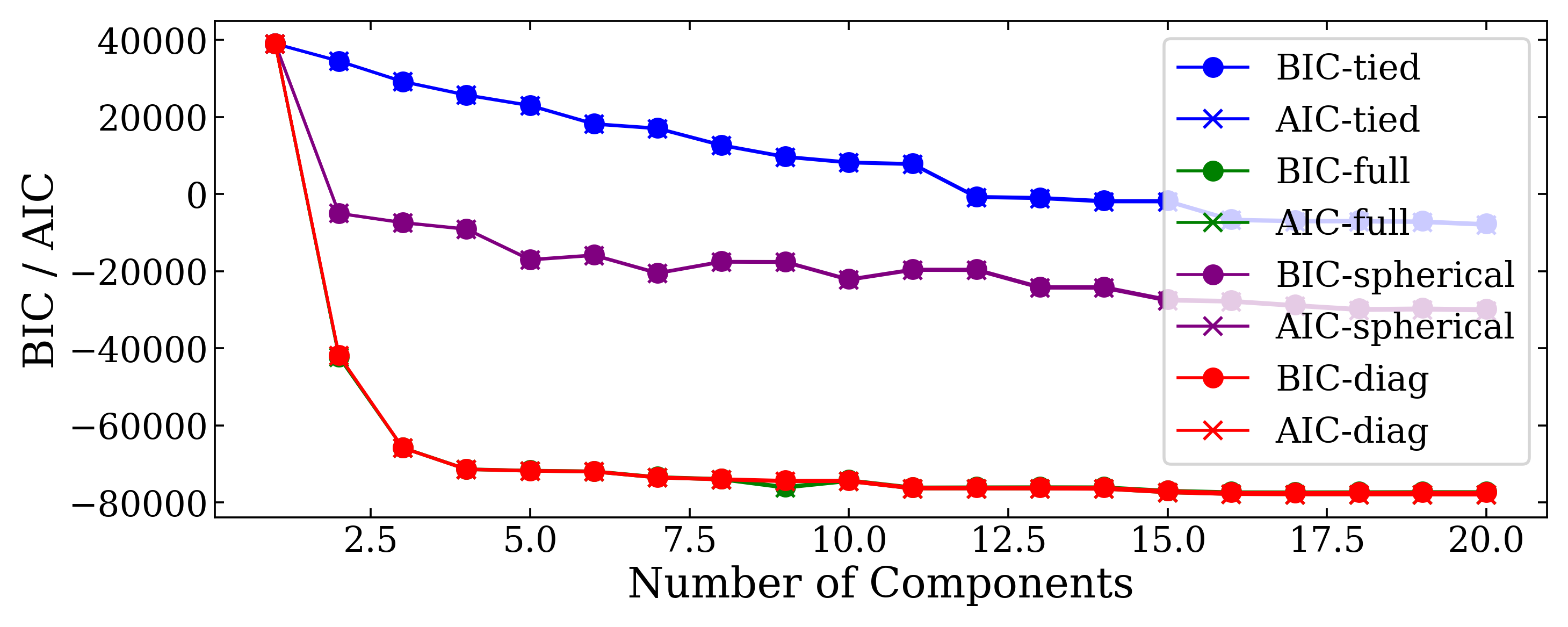}
        \caption{Beijing.}
        \label{fig:bic_aic_beijing}
    \end{subfigure}
    \caption{BIC and AIC values for GMM with varying numbers of clusters and different covariance types. Both subfigures share the same legend.}
    \label{fig:bic_aic}
\end{figure*}

Human beings, as inherently spatial creatures, intricately intertwine their activities with the surrounding geographical environments~\cite{lefebvre1991}. The organization of daily activities is significantly influenced by spatial factors, and residential decisions are often based on proximity to workplaces, schools, and essential amenities~\cite{cui_zhao_li_li_gong_deng_si_yan_dang_2024}. Similarly, workplaces are often selected for their accessibility and connectivity, while recreational areas are chosen for their appealing spatial qualities and ease of access~\cite{doi:10.1080/02673030500062335}.

Despite these general tendencies, the perception and navigation of geographical space vary significantly among individuals. Certain locations, such as homes, workplaces, and grocery stores, constitute the core of daily routines, frequented regularly for work, errands, and other obligations. These spaces seamlessly integrate into daily life, navigated with a habitual familiarity that underscores their importance. Conversely, leisure destinations, including parks, cinemas, and shopping malls, are sought for relaxation and recreational activities. Moreover, certain regions are encountered more casually, often during travel or exploration. These might encompass routes taken for vacations, spontaneous visits to new areas, or exploratory walks through different neighborhoods. Engaged with a sense of discovery and curiosity, these spaces are marked by less routine and more spontaneous interactions.

Understanding the connections between specific locations and individuals is essential for distinguishing predictable patterns from those that are uncertain and require further attention. In this study, we aim to capture and articulate the intricate relationships individuals maintain with their spatial environments.

The frequency and duration of visits to a location are robust indicators of its significance to an individual. To address this, we propose characterizing an individual's mobility trace by classifying their visits based on the number of days they visit and the duration of their stay at the considered PoIs.

We employ the Gaussian Mixture Model (GMM) to cluster PoIs at the user level based on visit frequency and dwell time. To ensure the effectiveness of our clustering, we determine the optimal number of clusters by computing the Bayesian Information Criterion (BIC)~\cite{https://doi.org/10.1002/wics.199} and the Akaike Information Criterion (AIC)~\cite{Bozdogan1987}. We explore a range of cluster numbers from 1 to 21 and evaluate four different covariance constraints: spherical, diagonal, tied, and full.

The BIC and AIC values for each combination of cluster count and covariance type are displayed in Figures~\ref{fig:bic_aic_singapore} and~\ref{fig:bic_aic_beijing}. These visuals help us understand the balance between model complexity and fit quality. In Figure~\ref{fig:bic_aic_singapore}, a clear elbow point is observable at 7 clusters, particularly for the tied covariance type. This finding is further supported by a zoomed-in inset, which also shows a distinct elbow at 7 clusters for the tied covariance type. Visually, the results of clustering with 7 clusters effectively capture the main structures in the data, providing meaningful and interpretable segmentations of PoIs characterization.

\begin{figure}[ht]
    \centering
    \includegraphics[width=8cm]{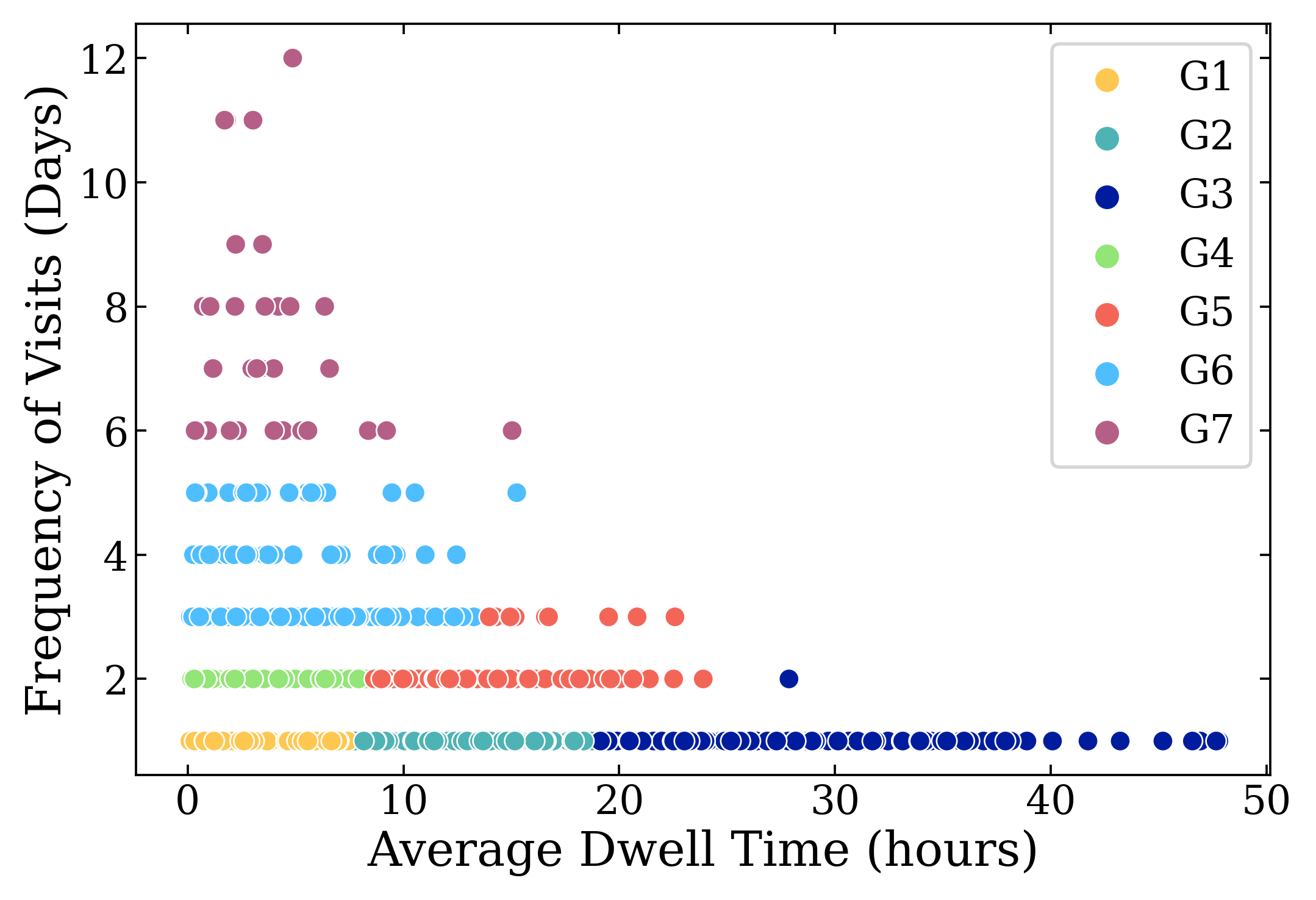}
    \caption{Visits characterization Beijing (GMM).}
    \label{fig:gmm_beijing}
\end{figure}

Following the determination of the optimal number of clusters, we apply the GMM clustering algorithm with seven components and a tied covariance type. We report the obtained results in Figures~\ref{fig:gmm_singapore} and~\ref{fig:gmm_beijing}. Despite differences in the dataset duration between Singapore and Geolife, we observe the formation of similar groups:

\begin{figure}[ht]
    \centering
    \includegraphics[width=8cm]{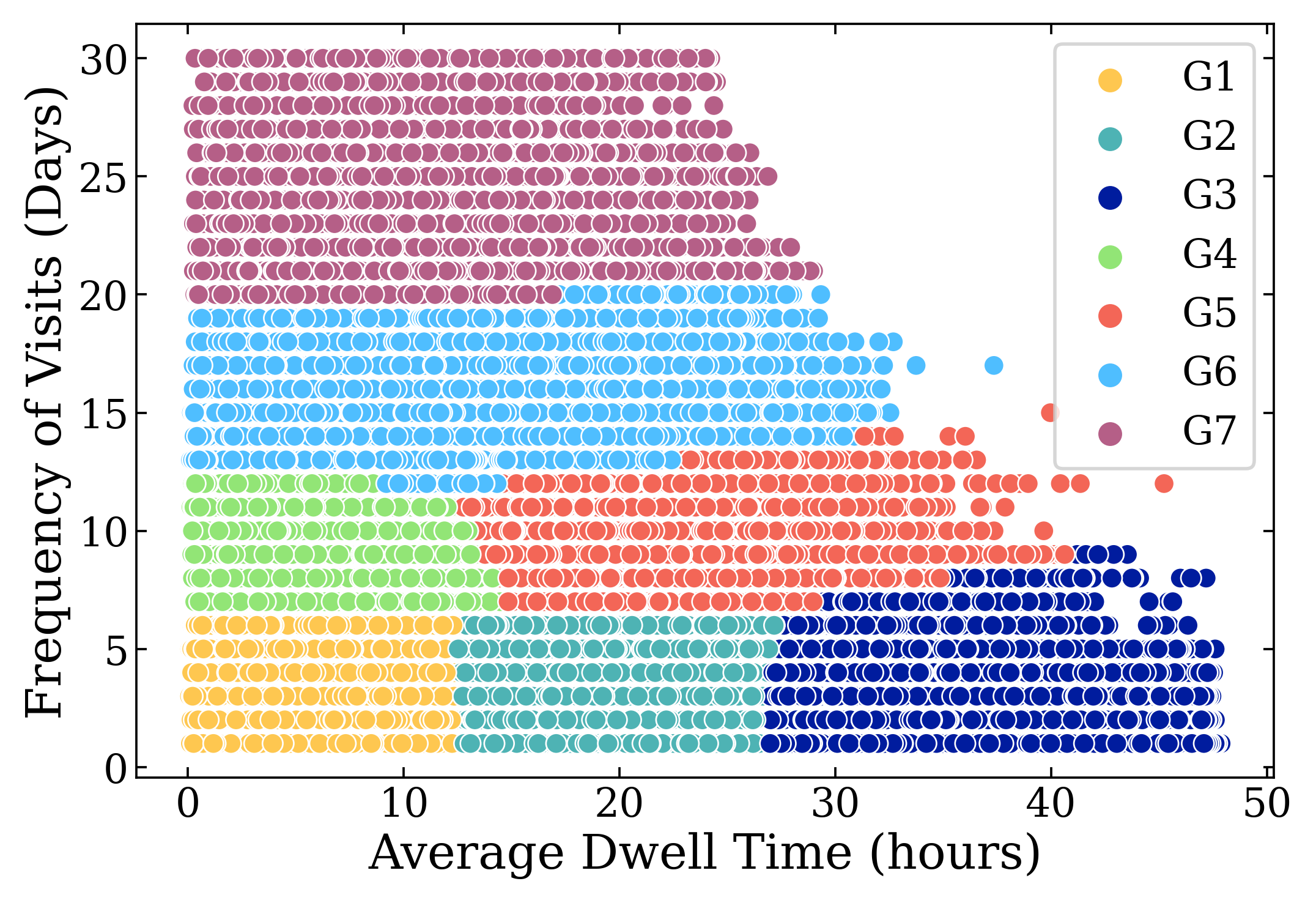}
    \caption{Visits characterization Singapore (GMM).}
    \label{fig:gmm_singapore}
\end{figure}

\begin{figure*}[!ht]
    \centering
    \begin{subfigure}[b]{0.3\textwidth}
        \centering
        \includegraphics[width=\textwidth]{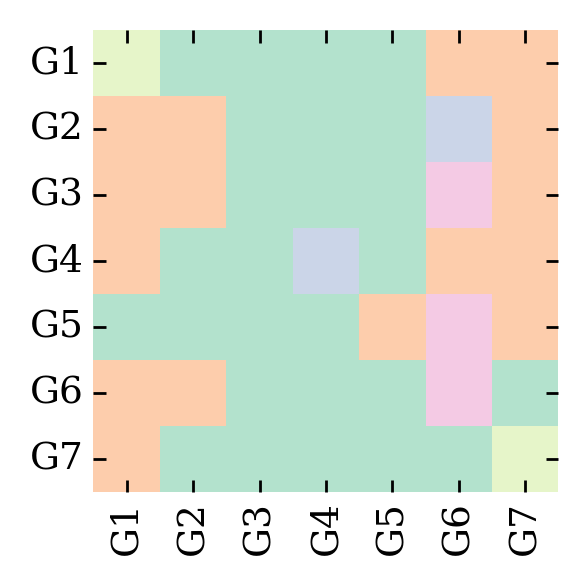}
        \caption{Visitation Pattern 1.}
        \label{fig:visitation_patterns_singapore_0}
    \end{subfigure}
    \hfill
    \begin{subfigure}[b]{0.3\textwidth}
        \centering
        \includegraphics[width=\textwidth]{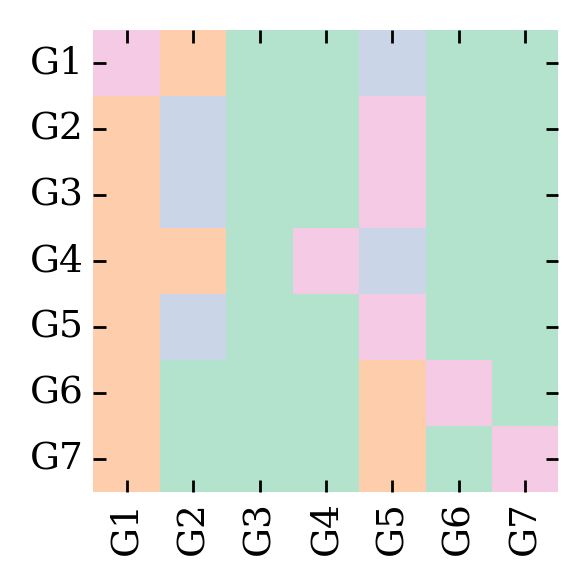}
        \caption{Visitation Pattern 2.}
        \label{fig:visitation_patterns_singapore_1}
    \end{subfigure}
    \hfill
    \begin{subfigure}[b]{0.38\textwidth}
        \centering
        \includegraphics[width=\textwidth]{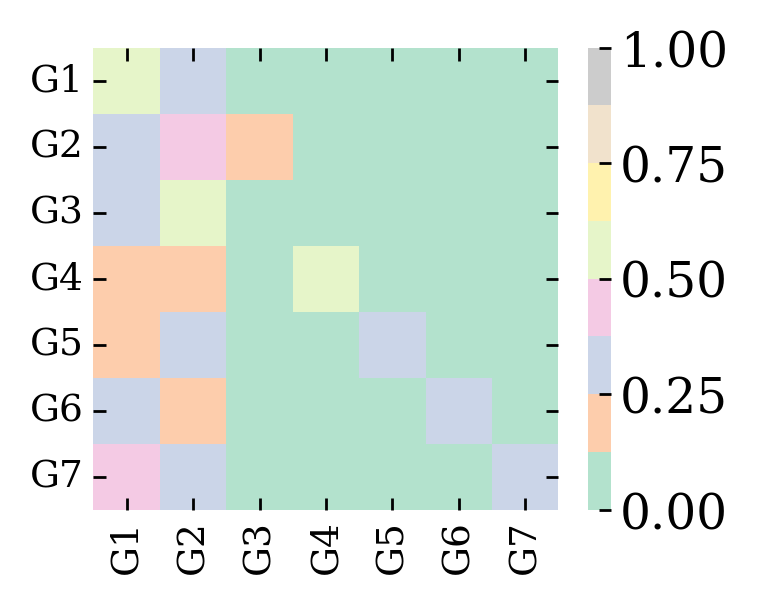}
        \caption{Visitation Pattern 3.}
        \label{fig:visitation_patterns_singapore_2}
    \end{subfigure}
    \caption{Singapore visitation patterns.}
    \label{fig:visitation_patterns_singapore}
\end{figure*}

\begin{figure*}[ht]
    \centering
    \begin{subfigure}[b]{0.3\textwidth}
        \centering
        \includegraphics[width=\textwidth]{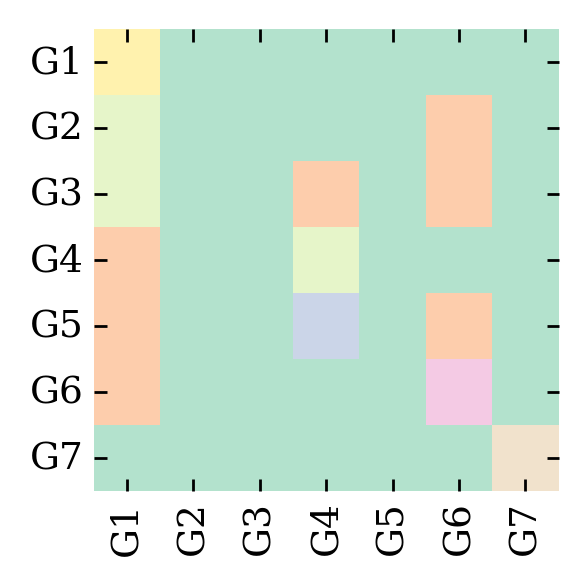}
        \caption{Visitation Pattern 1.}
        \label{fig:visitation_patterns_beijing_0}
    \end{subfigure}
    \hfill
    \begin{subfigure}[b]{0.3\textwidth}
        \centering
        \includegraphics[width=\textwidth]{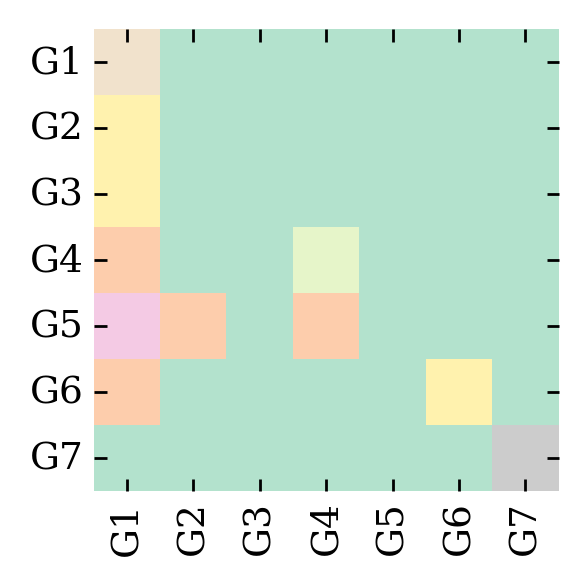}
        \caption{Visitation Pattern 2.}
        \label{fig:visitation_patterns_beijing_1}
    \end{subfigure}
    \hfill
    \begin{subfigure}[b]{0.38\textwidth}
        \centering
        \includegraphics[width=\textwidth]{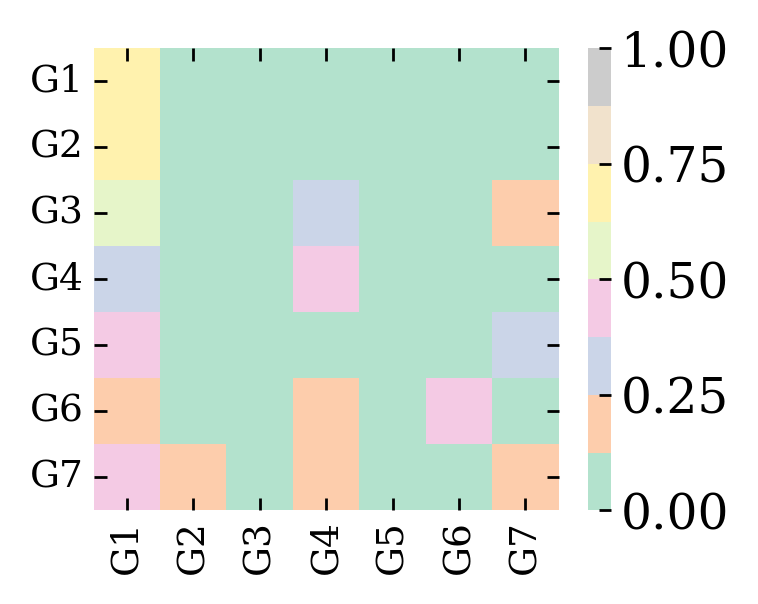}
        \caption{Visitation Pattern 3.}
        \label{fig:visitation_patterns_beijing_2}
    \end{subfigure}
    \caption{Beijing visitation patterns.}
    \label{fig:visitation_patterns_beijing}
\end{figure*}

\begin{itemize}
    \item \textbf{G1}: These are visits to locations that occur less frequently than the lower 20\% for both frequency and dwell distributions. These \textit{short-duration}, exploratory visits represent places individuals seldom go to, reflecting infrequent and sporadic mobility behavior.

    \item \textbf{G2}: Locations visited rarely (within the bottom 20\% of the frequency distribution) but where individuals spend extended time, typically between 10 hours and less than a full day. These locations often correspond to special events or sites of significant personal interest, representing \textit{long-duration}, rare explorations.

    \item \textbf{G3}: These visits are to locations that represent a change in the user's routine, often indicating a significant shift in their usual patterns of behavior. These locations typically have high dwell times exceeding a full day and varying visit frequencies, suggesting a \textit{temporary change in routine}.

    \item \textbf{G4}: These are \textit{casual} visits to locations where the frequency of visits falls within  20\% (16\%) and 44\% (33\%) for Singapore (Beijing) of the distribution, and the average dwell time is within the lower 20\%. While these locations are part of an individual's mobility life, they are not a regular part of their daily routine. Examples include a casually visited restaurant or pharmacy, where visits are brief and not frequent.

    \item \textbf{G5}: These visits occur with moderate frequency, generally between 20\% and 40\% of the frequency distribution, and involve spending a moderate to high amount of time. These locations hold more \textit{importance} than casual visits but are less frequent than routine ones.

    \item \textbf{G6}: These are visits to places that an individual frequents regularly, falling within the upper 60\% and 40\% of the frequency distribution. These locations are integral to the individual's daily \textit{routine} and typically have moderate dwell times.

    \item \textbf{G7}: These refer to \textit{anchored-like} locations where an individual spends a significant amount of time daily, falling within the upper 20\% of the frequency distribution. Such locations are characterized by frequent visits and prolonged dwell time, making it a central part of the individual's routine.
    
\end{itemize}

In summary, the seven visit categories capture distinct mobility patterns, ranging from exploratory to routine behaviors, characterized by variations in visit frequency and dwell time. These patterns offer insights into individuals' interactions with their environment, from sporadic visits to locations of personal significance to the establishment of routine, anchored places central to daily life.

\subsection{Visitation motifs}
Based on our proposed visitation classification, we identified the three dominant visitation patterns in both Singapore and Beijing, as shown in Figures~\ref{fig:visitation_patterns_singapore} and~\ref{fig:visitation_patterns_beijing}, respectively.

In Singapore (see Figures~\ref{fig:visitation_patterns_singapore}), emergent visitation behaviors provide key insights into individual mobility patterns. The first pattern (Figure~\ref{fig:visitation_patterns_singapore_0}) demonstrates that short-duration exploratory visits to \textit{G1} locations often lead to further exploratory activity, with occasional transitions to more stable, routine and anchored-like locations such as \textit{G6} and \textit{G7}. Visits to long-term exploratory sites, like \textit{G2} and \textit{G3}, frequently culminate in \textit{G6} visits, signifying a move toward more routine behavior. Individuals making brief, familiar visits to \textit{G4} typically continue with similar visit types, rarely shifting to exploratory or routine-altering activities. Following visits to \textit{G5} or \textit{G6}, individuals generally return to \textit{G6}, indicative of a stabilized routine. Meanwhile, visits to \textit{G7}—reflecting deep anchored routine—are predominantly followed by subsequent \textit{G7} visits, reinforcing a steady, habitual mobility patterns. This pattern is characterized by individuals who engage in short exploratory visits but tend to follow stable routine behaviors. In contrast, Figure~\ref{fig:visitation_patterns_singapore_1} highlights frequent transitions between exploratory-like visits and routine-like visits compared to the first pattern, such as \textit{G1,G2,G3}-to-\textit{G5} and \textit{G5}-to-\textit{G2}. This group also exhibits longer dwell times, regardless of whether their activities are exploratory or routine. Finally, Figure~\ref{fig:visitation_patterns_singapore_2} underscores a higher level of exploratory activity, both short- and long-term, with significant disruptions to routine (e.g., transitions to \textit{G3}). Visits to routine and anchored locations are still common for this subset, but the exploratory behavior leads to more dynamic mobility patterns.

In Beijing (see Figures~\ref{fig:visitation_patterns_beijing}), the first visitation pattern (Figure~\ref{fig:visitation_patterns_beijing_0}) reflects a distinct behavioral trend characterized by frequent yet brief visits to \textit{G1} locations, indicating exploratory activities. These visits are often complemented by casual, short-duration stops at familiar places, such as \textit{G4}, signaling a preference for familiar but flexible engagements. Moreover, individuals adhering to this pattern exhibit a pronounced tendency toward routine. This is evidenced by their consistent transitions to more stable, anchored locations, such as \textit{G6} and \textit{G7}, suggesting a balance between exploration and the need for stable, routine environments. The second visitation pattern (Figure~\ref{fig:visitation_patterns_beijing_1}) highlights a group with high levels of exploratory activities, both in short- and long-duration visits. These individuals demonstrate a propensity for breaking routine, engaging frequently in activities outside of their established patterns. Additionally, Figure~\ref{fig:visitation_patterns_beijing_1} depicts a marked tendency toward highly stable mobility patterns, as showed by frequent \textit{G7}-to-\textit{G7} transitions. This indicates a dual behavior where routine-breaking exploration coexists with a strong attachment to certain stable, anchored locations. Unlike the first pattern, where transitions to stable locations often occur directly after exploratory visits, this group tends to return to routine locations following visits to other routine sites. Lastly, the third visitation pattern (Figure~\ref{fig:visitation_patterns_beijing_2}) represents a subset of the population that primarily engages in exploratory visits, with only minimal routine behavior. This group appears to favor flexible and dynamic mobility, with limited engagement in consistent, anchored routines, distinguishing them as a highly exploratory subset within the population.

In both Singapore and Beijing, individuals balance exploratory and routine mobility behaviors, but their patterns diverge in key ways. In Singapore, transitions between exploratory and stable locations are more fluid, with individuals spending longer periods in exploratory visits (\textit{G2} and \textit{G3}). In contrast, mobility patterns in Beijing show a stronger attachment to routine, with a predominance of \textit{G7} visits, where individuals frequently return to stable, anchored locations after brief exploratory activities.

\subsection{Semantic patterns}
\begin{figure}[ht]
\centering
    \begin{subfigure}[b]{.45\textwidth}
        \centering
        \includegraphics[width=8.2cm]{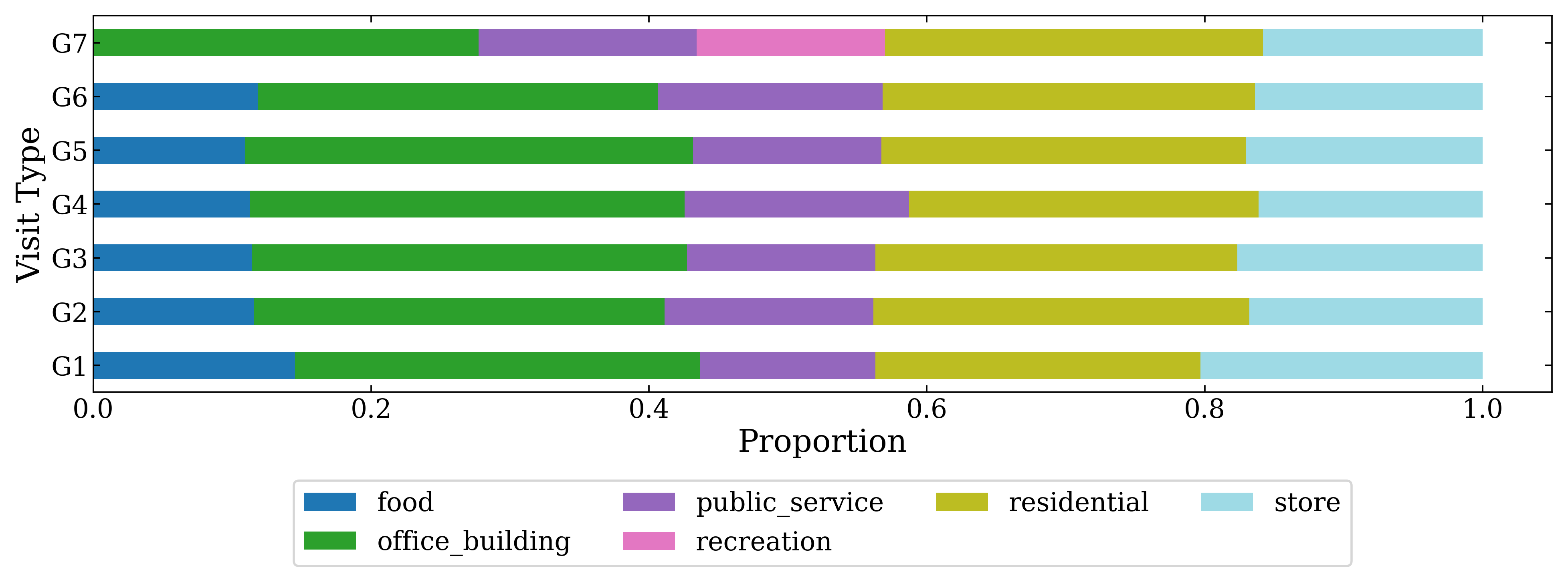}
        \caption{Singapore.}
        \label{fig:semantic_visits_singapore}
    \end{subfigure}
\hfill
    \begin{subfigure}[b]{.45\textwidth}
        \centering
    \includegraphics[width=8.2cm]{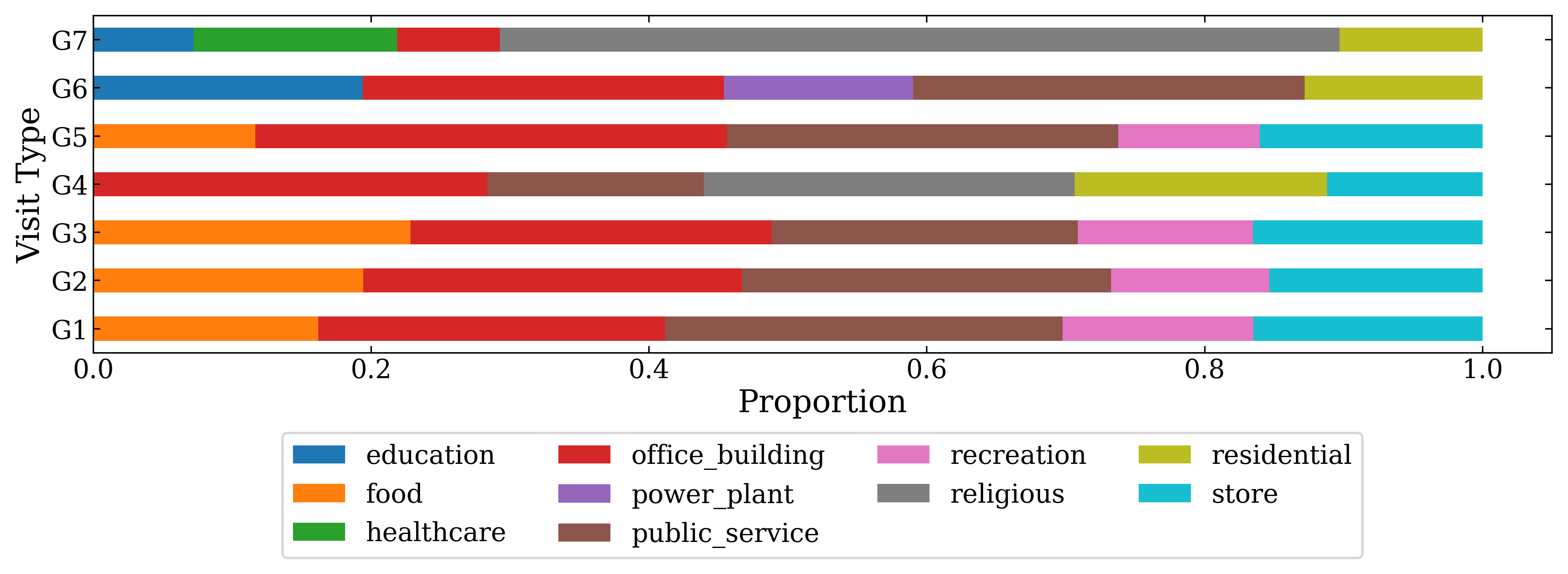}
    \caption{Beijing.}
    \label{fig:semantic_visits_beijing}
    \end{subfigure}
\caption{Top semantic for each type of visit.}
\label{fig:semantic_patterns}
\end{figure}

In Figure~\ref{fig:semantic_patterns}, we present the top five location types (semantics) visited across the different identified visitation categories. 

\vspace{-1mm}

In the case of Singapore (Figure~\ref{fig:semantic_visits_singapore}), frequently visited location types include `food', `public\_service', `residential', `store', and `office\_building', which are visited across all categories of visits, but in varying proportions. Particularly for \textit{G7} visits, the appearance of `recreation' as a top semantic category suggests a notable representation of workers employed in recreational areas within the Singapore dataset.

For Beijing (Figure~\ref{fig:semantic_patterns}), the semantic categories of `office\_building' and `public\_service' consistently appear almost across all visitation types, reflecting a mixture of regular employees and transient visitors. `Food' establishments are primarily visited during \textit{G1}, \textit{G2},  \textit{G3}, and \textit{G5} visits, indicating two distinct visitor profiles: leisure visitors, who are captured through \textit{G1} visits, and individuals working in the food industry, as inferred from \textit{G5} visits characterized by extended durations and frequent occurrences.

The presence of `education' in \textit{G7} and \textit{G6} visits likely reflects the mobility patterns of students and educational staff captured in the Geolife dataset. Similarly, the appearance of `power\_plant' in \textit{G6} visits suggests the inclusion of individuals employed at power plants. An additional semantic category of note is `religious', which appears in the top five for \textit{G7} and \textit{G4} visits, highlighting both full-time staff at religious institutions and casual attendees. `Residential' and `store' locations are among the top five for several visit categories. Lastly, `recreation' emerges in \textit{G5}, \textit{G3}, \textit{G2}, and \textit{G1} visits, signifying the presence of both workers and individuals frequenting recreational venues.

In both Singapore and Beijing, key location types like `public\_service', `residential', and `office\_building' dominate across visitation categories, with unique patterns emerging in specific contexts. Notable distinctions include recreational areas in Singapore's \textit{G7} visits and the presence of `food', `education', and power-related locations in Beijing, reflecting diverse visitor profiles.

\subsection{Temporal patterns}
\begin{figure}[ht]
\centering
\begin{subfigure}[b]{.45\textwidth}
    \centering
    \includegraphics[width=6cm]{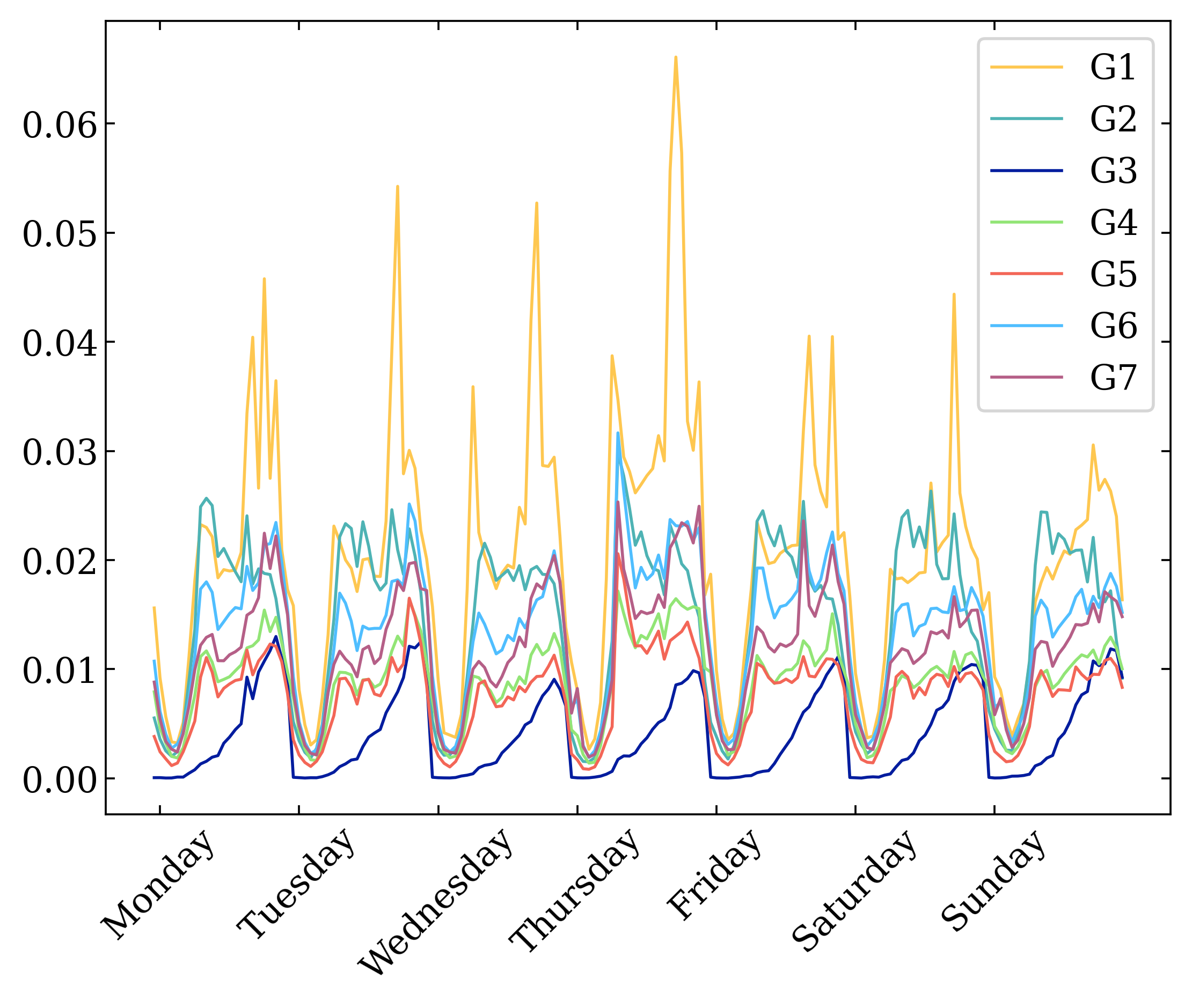}
    \caption{Singapore.}
    \label{fig:temporal_activity_singapore}
\end{subfigure}
\hfill
    \begin{subfigure}[b]{.45\textwidth}
    \centering
    \includegraphics[width=6cm]{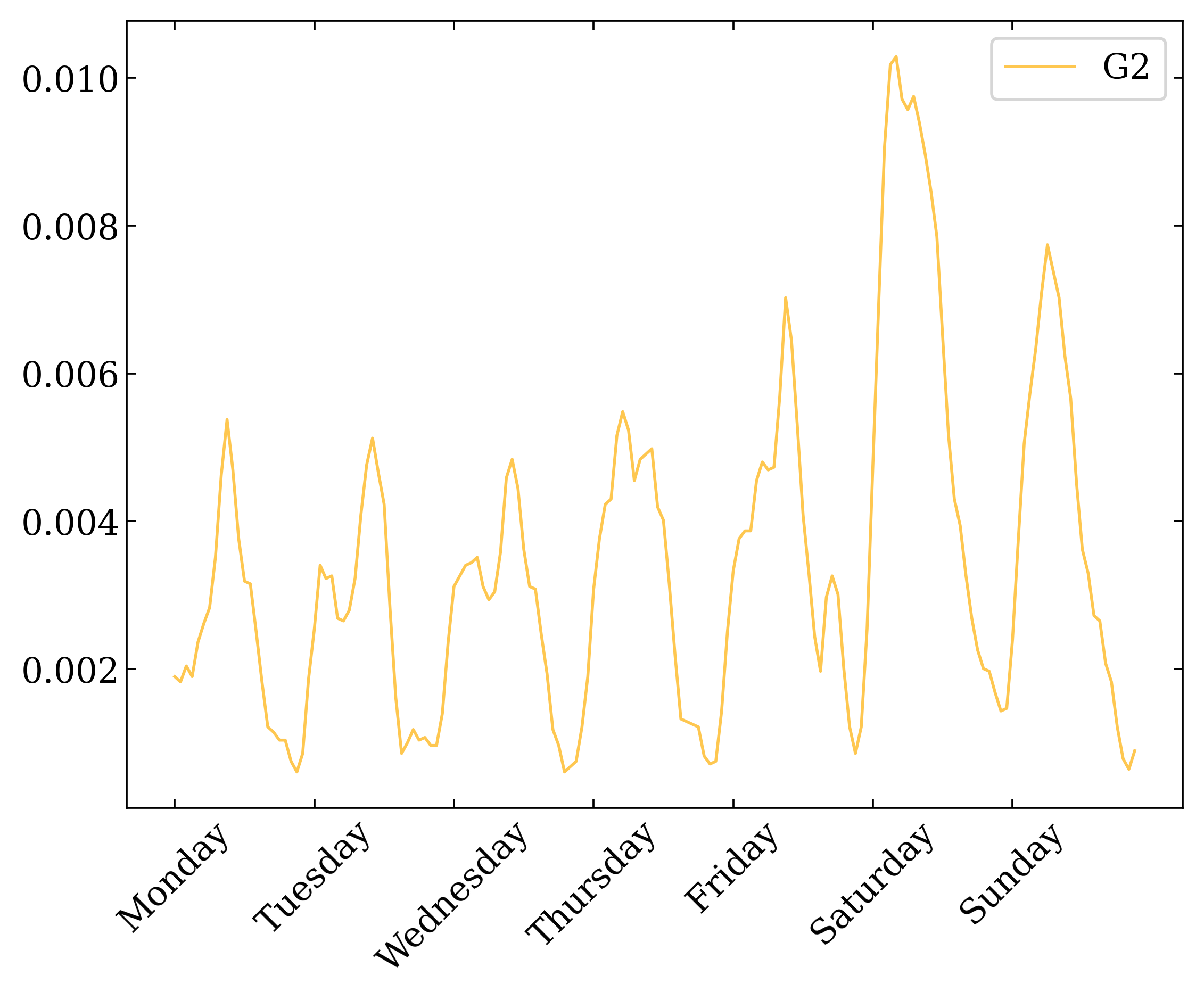}
    \caption{Beijing.}
    \label{fig:temporal_activity_beijing}
\end{subfigure}
    \caption{Temporal patterns.}
    \label{fig:temporal_patterns}
\end{figure}

In Figure~\ref{fig:temporal_patterns}, we present the weekly temporal patterns of different visit types, normalized across both weeks and users.

For Singapore (cf.~\ref{fig:temporal_activity_singapore}), \textit{G1} short-duration exploratory visits exhibit significant peaks late in the day and some early in the morning, making them the most prevalent type of visit. This pattern reflects the population's high tendency to explore new locations for short durations, particularly on Thursday evenings. In Singapore, Thursday evenings are commonly associated with social gatherings and after-work meetups, encouraging exploratory behavior. \textit{G2} visits remain stable throughout the week, with a slight increase on Saturdays when individuals have more time for extended explorations. Meanwhile, \textit{G3} visits, which signal routine changes, are consistent throughout the week, peaking in the late afternoon and evening, indicating shifts in nighttime activities. Visits to \textit{G5}, \textit{G6}, and \textit{G7} locations exhibit stability during the week, with a slight rise on Thursdays, followed by a decline over the weekend. %\textbf{}

In Figure~\ref{fig:temporal_activity_beijing}, we present the temporal patterns for \textit{G1} visits in Beijing. Unlike other visit types, which exhibit less consistent patterns due to the limited data available in the Geolife dataset, \textit{G1} visits show a more stable trend throughout the week. Notably, there is a significant increase in \textit{G1} visits during the weekend, particularly on Saturdays. This spike suggests that users have more leisure time to explore new locations, highlighting a clear pattern of increased exploration and activity during the weekend.

The overall skewness observed in visit patterns for Beijing is attributed to the constraints of the dataset, which impacts the accuracy and clarity of temporal dynamics for other visit types. The more extensive data available for Singapore enables a clearer characterization of temporal patterns, providing a more comprehensive understanding of visit behaviors.

\subsection{Spatial patterns}

\begin{figure}[ht]
\begin{subfigure}[b]{.45\textwidth}
    \centering
    \includegraphics[width=7cm]{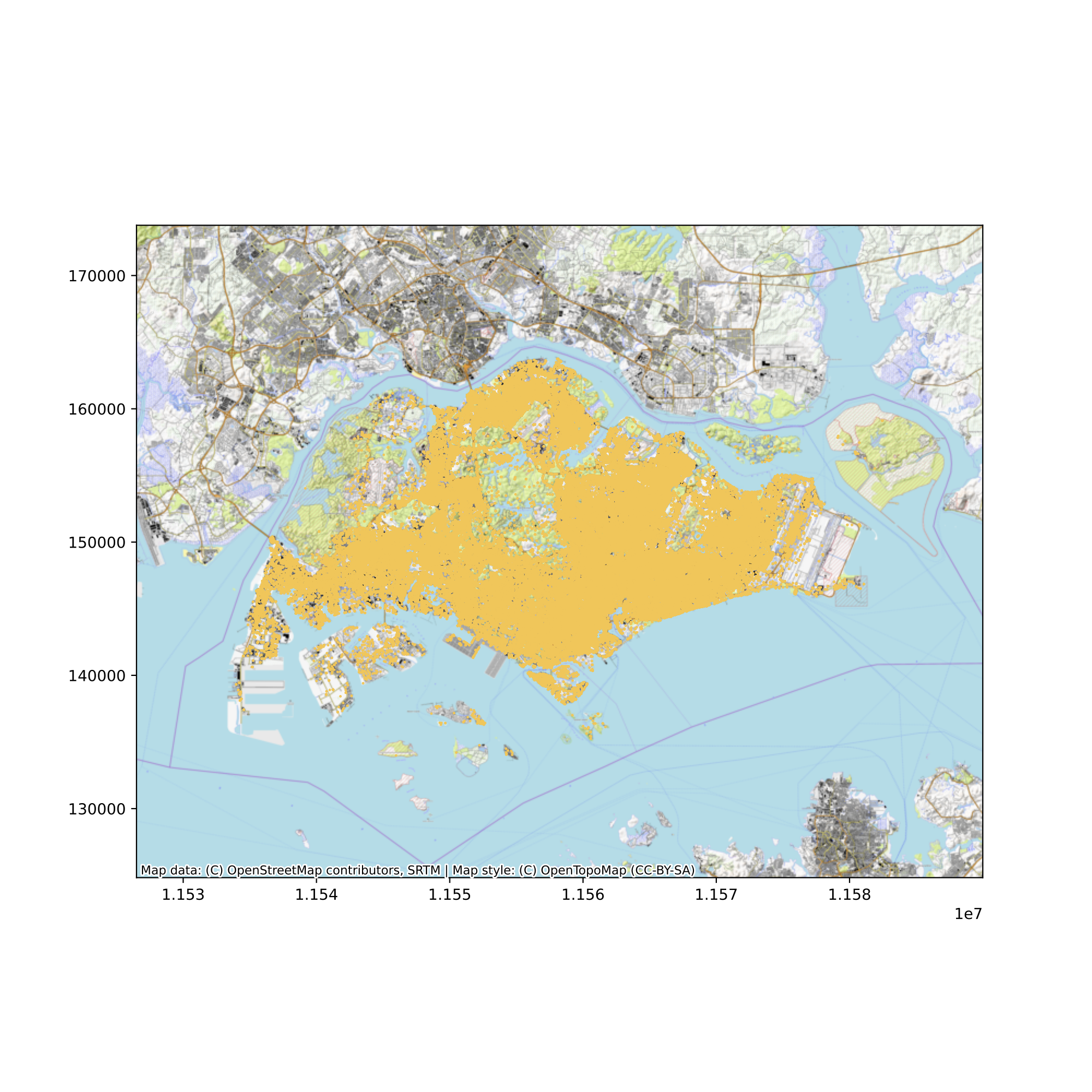}
    %\vspace{-2cm}
    \caption{G1.}
    \label{fig:spatial_occasional}
\end{subfigure}
\hfill
    %\vspace{-.5cm}
    \begin{subfigure}[b]{.45\textwidth}
    \centering
    \includegraphics[width=7cm]{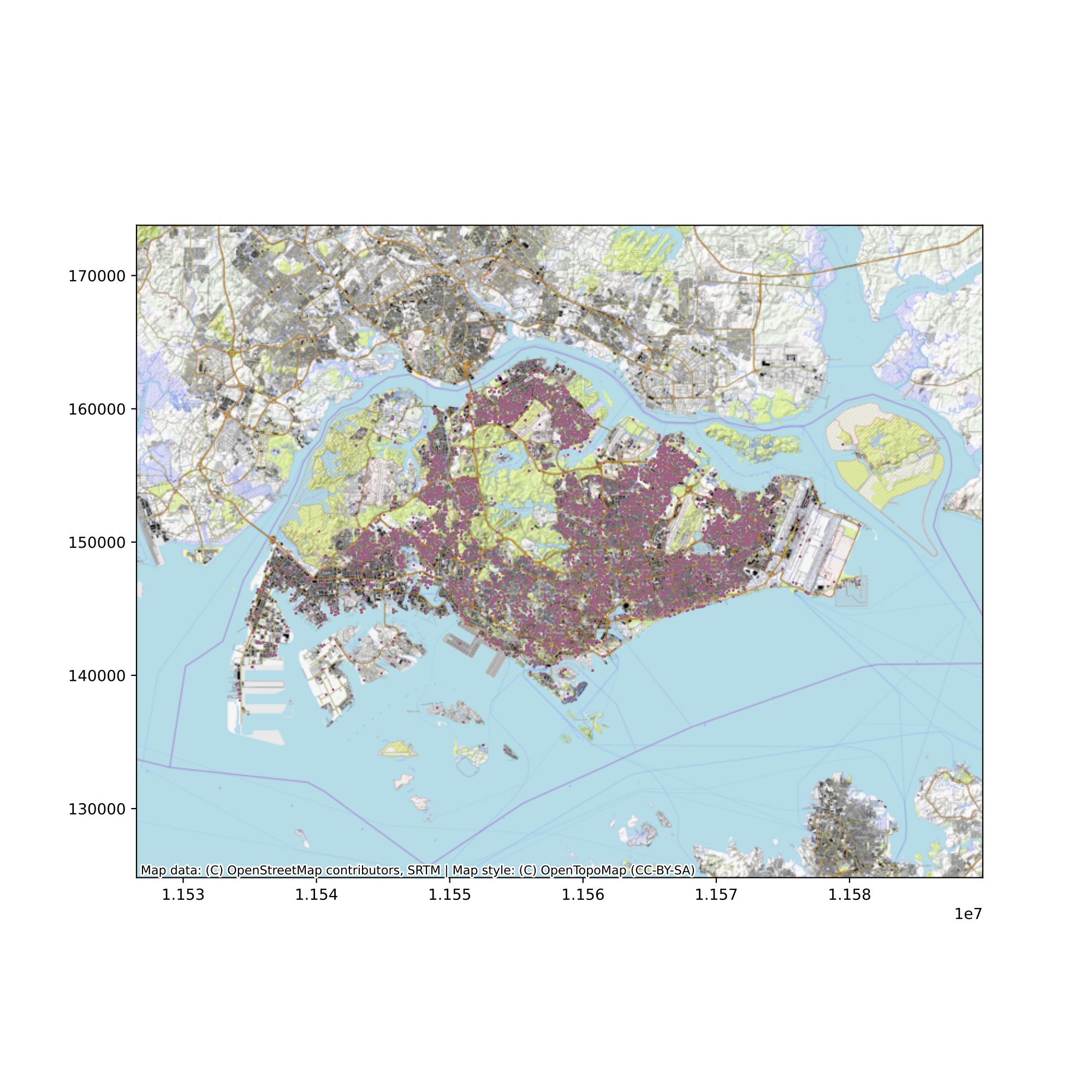}
    %\vspace{-1cm}
    \caption{G7.}
    \label{fig:spatial_anchors}
\end{subfigure}
\caption{Spatial exploitation (Singapore).}
    \label{fig:spatial_patterns}
\end{figure}

In Figure~\ref{fig:spatial_patterns}, we illustrate the spatial distribution of two contrasting visit types in Singapore based on our proposed classification: short-duration exploratory visits \textit{G1} and anchored-like visits \textit{G7}. \textit{G1} visits exhibit a broad spatial dispersion across the AOI, indicating a diverse range of locations frequented by users with varying patterns. This wide dispersion suggests that users with \textit{G1} visits engage in a variety of activities across diverse settings, reflecting a high level of mobility within the city. In contrast, \textit{G7} visits are notably concentrated around residential areas. This spatial concentration highlights the tendency for \textit{G7} visits to occur in more familiar, localized settings, reflecting their association with extended dwell times and frequent returns to home or central locations.

\begin{figure}[ht]
\centering
\begin{subfigure}[b]{.45\textwidth}
    \centering
    \includegraphics[width=5.5cm]{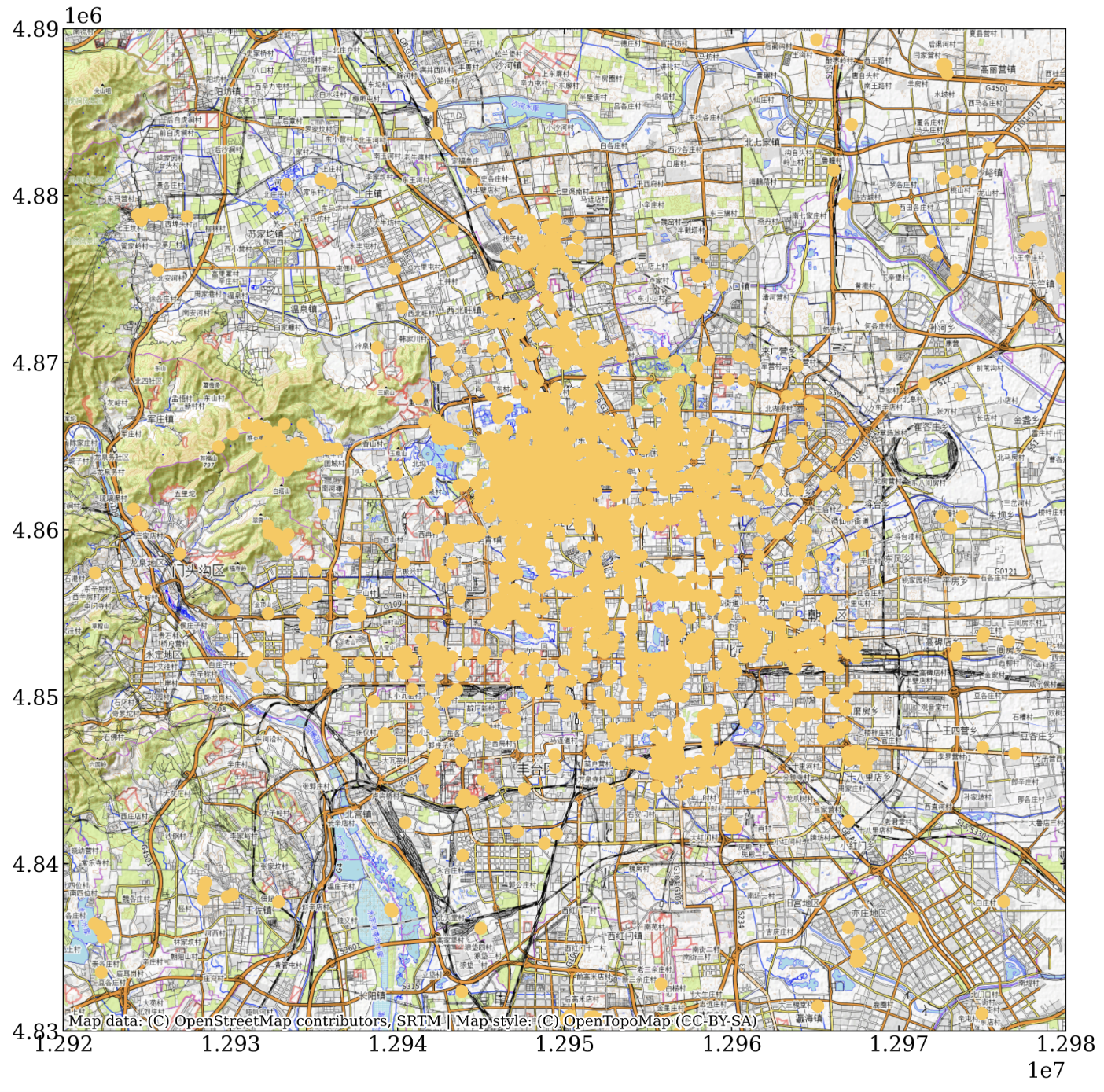}
    %\vspace{-2cm}
    \caption{G1.}
    \label{fig:spatial_occasional_beijing}
\end{subfigure}
\hfill
    %\vspace{-.5cm}
    \begin{subfigure}[b]{.45\textwidth}
    \centering
    \includegraphics[width=5.5cm]{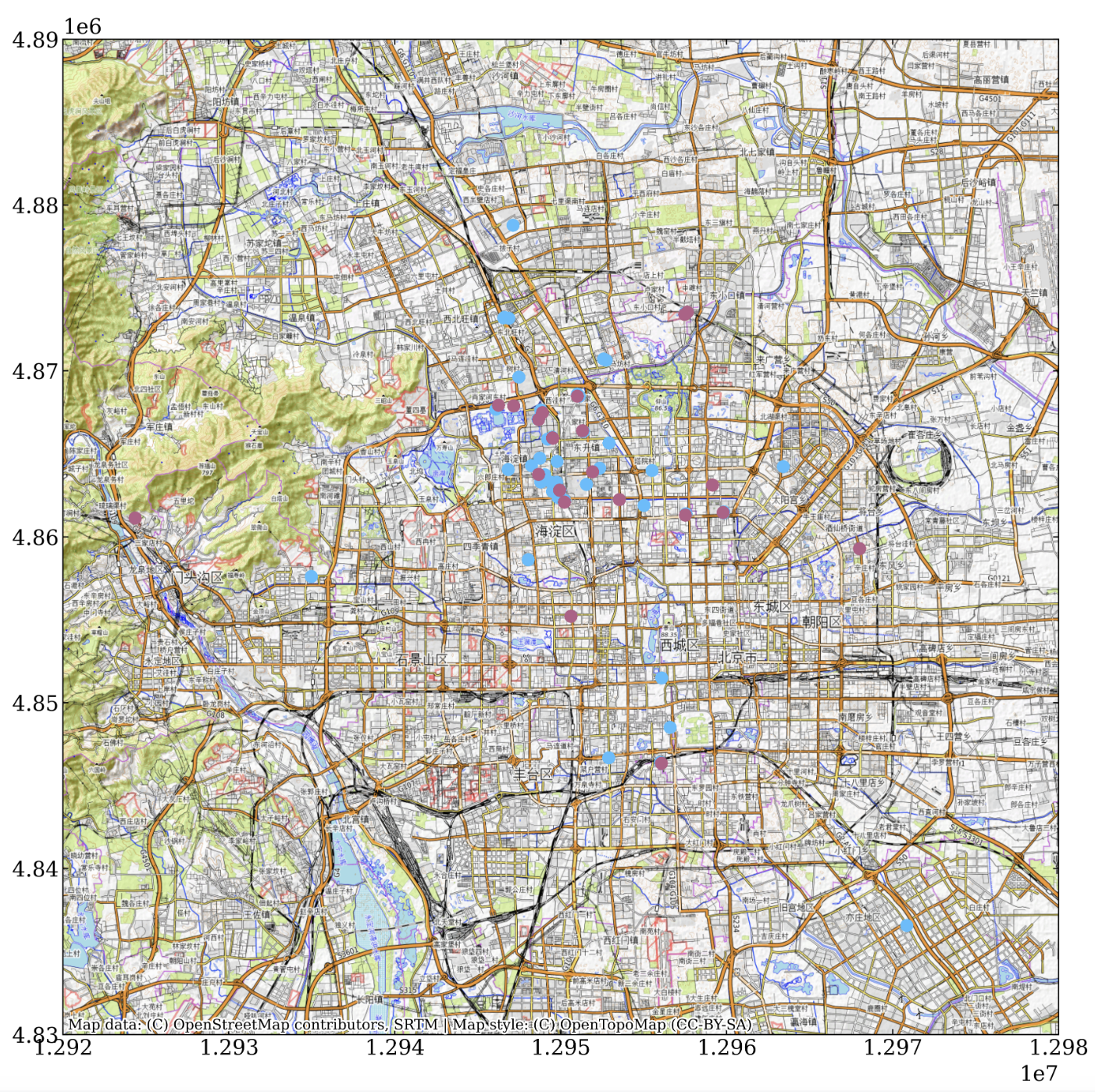}
    %\vspace{-1cm}
    \caption{G6-G7.}
    \label{fig:spatial_anchors_beijing}
\end{subfigure}
\caption{Spatial exploitation (Beijing).}
    \label{fig:spatial_patterns_beijing}
\end{figure}

Similarly, for Beijing, Figure~\ref{fig:spatial_occasional_beijing} illustrates the spatial distribution of \textit{G1} visits, while Figure~\ref{fig:spatial_anchors_beijing} depicts the spatial patterns of both \textit{G7} and \textit{G6} visits combined, given their relatively low frequencies. Consistent with observations in Singapore, \textit{G1} visits in Beijing are broadly dispersed across the city, indicating a wide range of locations and activities engaged in by users. In contrast, \textit{G7} and \textit{G6} visits exhibit notable concentration in the upper part of downtown, particularly around Peking University and the Haidian business district. The concentration of \textit{G7} and \textit{G6} visits in these areas suggests a high level of activity related to education and business. The clustering in these central, well-established locations reflects the individuals' regular engagement with these key urban hubs.

\vspace{-1mm}

\section{Related Works}
The analysis of human mobility is a focal point in various domains, including location-based services, urban planning, and geosocial marketing. Traditional mobility models, such as Markovian-based approaches, have been widely used to predict future locations based on past visitation patterns~\cite{Song_2010,9992252}. While effective in capturing frequent transitions, these models often overlook the semantic significance of different locations, leading to a limited understanding of the motivations behind human movement~\cite{10.1145/3423334.3431449}. Recent advancements have sought to enhance mobility modeling by incorporating temporal and spatial dimensions. For instance, models like the Time-Geo framework have introduced time-aware location predictions, improving the accuracy of movement forecasts by considering the temporal aspects of visits~\cite{Jiang_2016}. Also, spatial clustering techniques have been employed to group similar locations, aiming to identify routine visits more effectively. However, these approaches often treat locations homogeneously, failing to distinguish between various types of visits that carry different meanings for individuals~\cite{9992252,pappalardo2015returners,10.1145/3397536.3422248}.

Several works have attempted to integrate semantic data into mobility models. For example, SemanticTraj utilized semantic annotations to classify trajectories, offering a more refined understanding of movement patterns~\cite{al2016semantictraj}. Similarly, the PlaceRank model introduced the concept of location importance, ranking locations based on visitation frequency and duration~\cite{WANG2015335}. Despite these advancements, the existing models typically focus on either spatial or semantic aspects in isolation, and they often lack a comprehensive framework that integrates semantic, spatial, and temporal dimensions~\cite{9992252}.

This paper addresses these gaps by introducing the a new method, which uniquely characterizes visits into seven distinct categories according to their degrees of explorations and routine. Unlike previous approaches, the proposed method leverages a multi-dimensional analysis that accounts for the semantic significance, spatial distribution, and temporal patterns of visits, providing a holistic view of human mobility.

In summary, while there has been substantial progress in the field of mobility modeling, the existing methods fall short of capturing the complex and multi-faceted nature of human-environment interactions. The visits characterization model bridges this gap by offering a more nuanced characterization of visits, paving the way for improved applications in location-based services, urban planning, and personalized recommendations.

%external .bib file (using bibtex).

%\begin{thebibliography}{1}
\bibliographystyle{unsrt}  
\bibliography{ref}  %%% Remove comment to use the 

\begin{thebibliography}{10}

\bibitem{agoda_2019}
68\% have visited up to 10 countries: Agoda.com study.

\bibitem{unwto_2023}
World tourism barometer.

\bibitem{barbosa2018human}
Hugo Barbosa, Marc Barthelemy, Gourab Ghoshal, Charlotte~R James, Maxime Lenormand, Thomas Louail, Ronaldo Menezes, Jos{\'e}~J Ramasco, Filippo Simini, and Marcello Tomasini.
\newblock Human mobility: Models and applications.
\newblock {\em Physics Reports}, 734:1--74, 2018.

\bibitem{pappalardo2023future}
Luca Pappalardo, Ed~Manley, Vedran Sekara, and Laura Alessandretti.
\newblock Future directions in human mobility science.
\newblock {\em Nature computational science}, 3(7):588--600, 2023.

\bibitem{10.1145/3240323.3240361}
Zhu Sun, Jie Yang, Jie Zhang, Alessandro Bozzon, Long-Kai Huang, and Chi Xu.
\newblock Recurrent knowledge graph embedding for effective recommendation.
\newblock In {\em Proceedings of the 12th ACM Conference on Recommender Systems}, RecSys '18, page 297–305, New York, NY, USA, 2018. Association for Computing Machinery.

\bibitem{10.1145/3494993}
Huandong Wang, Qiaohong Yu, Yu~Liu, Depeng Jin, and Yong Li.
\newblock Spatio-temporal urban knowledge graph enabled mobility prediction.
\newblock {\em Proc. ACM Interact. Mob. Wearable Ubiquitous Technol.}, 5(4), 12 2022.

\bibitem{cuttone2018understanding}
Andrea Cuttone, Sune Lehmann, and Marta~C Gonz{\'a}lez.
\newblock Understanding predictability and exploration in human mobility.
\newblock {\em EPJ Data Science}, 7:1--17, 2018.

\bibitem{Huandong_2022}
Huandong Wang, Qiaohong Yu, Yu~Liu, Depeng Jin, and Yong Li.
\newblock Spatio-temporal urban knowledge graph enabled mobility prediction.
\newblock {\em Proc. ACM Interact. Mob. Wearable Ubiquitous Technol.}, 5(4), 12 2022.

\bibitem{pappalardo2024survey}
Luca Pappalardo, Emanuele Ferragina, Salvatore Citraro, Giuliano Cornacchia, Mirco Nanni, Giulio Rossetti, Gizem Gezici, Fosca Giannotti, Margherita Lalli, Daniele Gambetta, et~al.
\newblock A survey on the impact of ai-based recommenders on human behaviours: methodologies, outcomes and future directions.
\newblock {\em arXiv preprint arXiv:2407.01630}, 2024.

\bibitem{gonzalez2008understanding}
Marta~C Gonzalez, Cesar~A Hidalgo, and Albert-Laszlo Barabasi.
\newblock Understanding individual human mobility patterns.
\newblock {\em nature}, 453(7196):779--782, 2008.

\bibitem{song2010modelling}
Chaoming Song, Tal Koren, Pu~Wang, and Albert-L{\'a}szl{\'o} Barab{\'a}si.
\newblock Modelling the scaling properties of human mobility.
\newblock {\em Nature physics}, 6(10):818--823, 2010.

\bibitem{luca2021survey}
Massimiliano Luca, Gianni Barlacchi, Bruno Lepri, and Luca Pappalardo.
\newblock A survey on deep learning for human mobility.
\newblock {\em ACM Computing Surveys (CSUR)}, 55(1):1--44, 2021.

\bibitem{10.1145/3677019}
Qiaohong Yu, Huandong Wang, Yu~Liu, Depeng Jin, Yong Li, Lin Zhu, and Junlan Feng.
\newblock Mobility prediction via rule-enhanced knowledge graph.
\newblock {\em ACM Trans. Knowl. Discov. Data}, 7 2024.
\newblock Just Accepted.

\bibitem{alessandretti2020scales}
Laura Alessandretti, Ulf Aslak, and Sune Lehmann.
\newblock The scales of human mobility.
\newblock {\em Nature}, 587(7834):402--407, 2020.

\bibitem{10.1145/3356991.3365474}
Rachel Palumbo, Laura Thompson, and Gautam Thakur.
\newblock Sonet: a semantic ontological network graph for managing points of interest data heterogeneity.
\newblock In {\em Proceedings of the 3rd ACM SIGSPATIAL International Workshop on Geospatial Humanities}, GeoHumanities '19, New York, NY, USA, 2019. Association for Computing Machinery.

\bibitem{osti_2000381}
Junchuan Fan, Joseph Bentley, and Gautam~Malviya Thakur.
\newblock Sonet++: A knowledge graph of geographic categories based on osm tag representation.
\newblock {\em ORNL}, 9 2023.

\bibitem{thakur2015planetsense}
Gautam~S Thakur, Budhendra~L Bhaduri, Jesse~O Piburn, Kelly~M Sims, Robert~N Stewart, and Marie~L Urban.
\newblock Planetsense: a real-time streaming and spatio-temporal analytics platform for gathering geo-spatial intelligence from open source data.
\newblock In {\em Proceedings of the 23rd SIGSPATIAL International Conference on Advances in Geographic Information Systems}, pages 1--4, 2015.

\bibitem{lefebvre1991}
Henri Lefebvre.
\newblock {\em The Production of Space}.
\newblock Blackwell, Cambridge, MA, 1991.
\newblock Original work published 1974.

\bibitem{cui_zhao_li_li_gong_deng_si_yan_dang_2024}
Yanzhe Cui, Pengjun Zhao, Ling Li, Juan Li, Mingyuan Gong, Yiling Deng, Zihuang Si, Shuaichen Yan, and Xuewei Dang.
\newblock A new model for residential location choice using residential trajectory data.
\newblock {\em Humanities and Social Sciences Communications}, 11(1), 2024.

\bibitem{doi:10.1080/02673030500062335}
Mark W.~Horner Tae-Kyung~Kim and Robert~W. Marans.
\newblock Life cycle and environmental factors in selecting residential and job locations.
\newblock {\em Housing Studies}, 20(3):457--473, 2005.

\bibitem{https://doi.org/10.1002/wics.199}
Andrew~A. Neath and Joseph~E. Cavanaugh.
\newblock The bayesian information criterion: background, derivation, and applications.
\newblock {\em WIREs Computational Statistics}, 4(2):199--203, 2012.

\bibitem{Bozdogan1987}
Hamparsum Bozdogan.
\newblock Model selection and akaike's information criterion (aic): The general theory and its analytical extensions.
\newblock {\em Psychometrika}, 52:345--370, 9 1987.

\bibitem{Song_2010}
Chaoming Song, Tal Koren, Pu~Wang, and Albert-L{\'a}szl{\'o} Barab{\'a}si.
\newblock Modelling the scaling properties of human mobility.
\newblock {\em Nature physics}, 6(10):818--823, 2010.

\bibitem{9992252}
Licia Amichi, Aline~Viana Carneiro, Mark Crovella, and Antonio Loureiro.
\newblock Revealing an inherently limiting factor in human mobility prediction.
\newblock {\em IEEE Transactions on Emerging Topics in Computing}, 11(3):635--649, 2023.

\bibitem{10.1145/3423334.3431449}
Gautam Thakur and Olivera Kotevska.
\newblock Activity characterization for modeling behavioral-driven human mobility in platial networks.
\newblock In {\em Proceedings of the 4th ACM SIGSPATIAL Workshop on Location-Based Recommendations, Geosocial Networks, and Geoadvertising}, LocalRec'20, New York, NY, USA, 2020. Association for Computing Machinery.

\bibitem{Jiang_2016}
Shan Jiang, Yingxiang Yang, Siddharth Gupta, Daniele Veneziano, Shounak Athavale, and Marta~C. González.
\newblock The timegeo modeling framework for urban mobility without travel surveys.
\newblock {\em Proceedings of the National Academy of Sciences}, 113(37):E5370--E5378, 2016.

\bibitem{pappalardo2015returners}
Luca Pappalardo, Filippo Simini, Salvatore Rinzivillo, Dino Pedreschi, Fosca Giannotti, and Albert-L{\'a}szl{\'o} Barab{\'a}si.
\newblock Returners and explorers dichotomy in human mobility.
\newblock {\em Nature communications}, 6(1):8166, 2015.

\bibitem{10.1145/3397536.3422248}
Licia Amichi, Aline~Carneiro Viana, Mark Crovella, and Antonio~A.F. Loureiro.
\newblock Understanding individuals' proclivity for novelty seeking.
\newblock In {\em Proceedings of the 28th International Conference on Advances in Geographic Information Systems}, SIGSPATIAL '20, page 314–324, New York, NY, USA, 2020. Association for Computing Machinery.

\bibitem{al2016semantictraj}
Shamal Al-Dohuki, Yingyu Wu, Farah Kamw, Jing Yang, Xin Li, Ye~Zhao, Xinyue Ye, Wei Chen, Chao Ma, and Fei Wang.
\newblock Semantictraj: A new approach to interacting with massive taxi trajectories.
\newblock {\em IEEE transactions on visualization and computer graphics}, 23(1):11--20, 2016.

\bibitem{WANG2015335}
Guihua Wang, Yuanguang Zhong, Chung-Piaw Teo, and Qizhang Liu.
\newblock Flow-based accessibility measurement: The place rank approach.
\newblock {\em Transportation Research Part C: Emerging Technologies}, 56:335--345, 2015.

\end{thebibliography}
%\end{thebibliography}
%\addbibresource{ref.bib}
%\printbibliography

\end{document}